\documentclass[lettersize,journal]{IEEEtran}
\usepackage{xcolor}
\usepackage{colortbl}
\usepackage{tcolorbox}
\usepackage{mdframed}
\usepackage{listings}
\usepackage{makecell}
\usepackage{multirow}
\usepackage{array}
\usepackage{cite} %
\usepackage{amsmath,amsfonts}

\usepackage{amssymb}
\usepackage{hyperref} 
\usepackage{algorithm}
\usepackage{algorithmic}
\usepackage{graphicx}
\usepackage{fbox}
\usepackage[caption=false,font=normalsize,labelfont=sf,textfont=sf]{subfig}
\usepackage{textcomp}
\usepackage{minibox}
\usepackage{balance}
\usepackage{stfloats}
\usepackage{verbatim}
\usepackage{enumitem}
\usepackage{xspace}
\usepackage[T1]{fontenc}
\usepackage{booktabs}
\usepackage{threeparttable}
\usepackage[]{collab}
\usepackage{url}
\usepackage{soul}

\usepackage[utf8]{inputenc}
\usepackage{color}
\usepackage{calligra}
\usepackage[normalem]{ulem}
\usepackage{indentfirst}
\usepackage{nicematrix}
\usepackage{fontawesome5}
\usepackage{pifont}
\usepackage{wrapfig}

\def\BibTeX{{\rm B\kern-.05em{\sc i\kern-.025em b}\kern-.08em
    T\kern-.1667em\lower.7ex\hbox{E}\kern-.125emX}}

\newcommand\task{LogAD\xspace}
\newcommand\nm{CodeAD\xspace}

\newcommand{\ie}{{i.e.},\xspace}
\newcommand{\eg}{{e.g.},\xspace}

\definecolor{ballblue}{rgb}{0.13, 0.67, 0.8}
\definecolor{jcpink}{RGB}{255, 0, 96}
\collabAuthor{jy}{ballblue}{Jinyang Liu}
\collabAuthor{jc}{jcpink}{JC}
\collabAuthor{jj}{purple}{Junjie}
\collabAuthor{mh}{brown}{Minghua}

\definecolor{mygreen}{HTML}{AFCFA5}
\newcounter{summary}
\newcommand{\summary}[1]{%
  \refstepcounter{summary}%
  \begin{mdframed}[linecolor=gray!25,roundcorner=12pt,
                   backgroundcolor=mygreen!20,linewidth=3pt,
                   innerleftmargin=2pt, leftmargin=0cm,
                   rightmargin=0cm, topline=false,
                   bottomline=false, rightline=false, leftline=false]
    #1%
  \end{mdframed}%
}

\definecolor{mygreen}{HTML}{AFCFA5}
\newcounter{implication}

\definecolor{myyellow}{HTML}{FFF2CC}
\newcounter{finding}

\begin{document}

\title{\nm: Synthesize Code of Rules for Log-based Anomaly Detection with LLMs}

\author{Junjie Huang\IEEEauthorrefmark{1}, Minghua He\IEEEauthorrefmark{2}, Jinyang Liu\IEEEauthorrefmark{1}, Yintong Huo\IEEEauthorrefmark{3}, Domenico Bianculli\IEEEauthorrefmark{4}, Michael R. Lyu\IEEEauthorrefmark{1}\\
\IEEEauthorblockA{
\IEEEauthorrefmark{1}The Chinese University of Hong Kong, Hong Kong SAR, \{jjhuang23, jyliu, lyu\}@cse.cuhk.edu.hk\\
\IEEEauthorrefmark{2}Peking University, Beijing, China, hemh2120@stu.pku.edu.cn\\
\IEEEauthorrefmark{3}Singapore Management University, Singapore, ythuo@smu.edu.sg\\
\IEEEauthorrefmark{4}University of Luxembourg, Luxembourg, domenico.bianculli@uni.lu\\
}
}

\markboth{Journal of \LaTeX\ Class Files,~Vol.~14, No.~8, August~2021}%
{Shell \MakeLowercase{\textit{et al.}}: A Sample Article Using IEEEtran.cls for IEEE Journals}

\maketitle

\begin{abstract}

Log-based anomaly detection (\task) is critical for maintaining the reliability and availability of large-scale online service systems.
While machine learning, deep learning, and large language models (LLMs)-based methods have advanced the \task, they often suffer from limited interpretability, high inference costs, and extensive preprocessing requirements, limiting their practicality for real-time, high-volume log analysis. In contrast, rule-based systems offer efficiency and transparency, but require significant manual effort and are difficult to scale across diverse and evolving environments.
In this paper, We present \nm, a novel framework that automatically synthesizes lightweight Python rule functions for LogAD using LLMs. 
\nm introduces a hierarchical clustering and anchor-grounded sampling strategy to construct representative contrastive log windows, enabling LLMs to discern discriminative anomaly patterns. To ensure robustness and generalizability, \nm employs an agentic workflow that iteratively generates, tests, repairs, and refines the rules until it meets correctness and abstraction requirements. The synthesized rules are interpretable, lightweight, and directly executable on raw logs, supporting efficient and transparent online anomaly detection.
Our comprehensive experiments on three public datasets (BGL, Hadoop, Thunderbird) demonstrate that \nm achieves an average absolute improvement of 3.6\% F1 score over the state-of-the-art baselines, while processing large datasets up to 4$\times$ faster and at a fraction of the cost (total LLM invocation cost under 4 USD per dataset).
These results highlight \nm as a practical and scalable solution for online monitoring systems, enabling interpretable, efficient, and automated \task in real-world environment.

\end{abstract}


\section{Introduction}

\IEEEPARstart{I}{n} recent years, IT companies started to deploy their applications and services to cloud platforms, providing 24x7 online services to hundreds of millions of users worldwide~\cite{chen2020icm}.
Ensuring the reliability and availability of online service systems are crucial for service providers, as a short period of interruption or performance degradation can lead to significant losses in user satisfaction and  revenue~\cite{zhang2025aiopssurvey, aws_news}.
To minimize negative impacts and safeguard reliability, a crucial step is the timely and accurate detection of anomalous system behaviors, \ie anomaly detection~\cite{soldani2022ADRCAsurvey}.
Logs are an important source in software systems that records software events during runtime~\cite{he2021logsurvey}. Analyzing logs is vital for software maintenance, and anomaly detection plays a key role in identifying system errors and potential risks~\cite{xu2009detecting, soldani2022ADRCAsurvey, ma2025practitioners}.

Over the years, many methods have been proposed for automatic log-based anomaly detection (hereafter shortened to \task). 
Machine learning (ML) and deep learning (DL) based methods train  models (\eg SVM~\cite{liang2007ibmSVM}, LSTM\cite{du2017deeplog}, transformers~\cite{guo2024logformer}) to classify a group of logs (often called a log window) based on extracted features. 
However, their \textit{efficiency} and \textit{interpretability} are limited. 
First, DL models require large-scale matrix operations and specialized hardware (\eg GPUs), making them unsuitable for real-time, large-scale log processing~\cite{yu2024DLorML}.
When processing, feature extraction (\eg log parsing to collect templates) is time-consuming, especially for complex logs. For instance, parsing 3 million logs in the Thunderbird dataset with Drain takes 38 minutes (\S\ref{sec:expr-rq4-time})~\cite{le2022DeepLogADStudy, ali2025mlLOD}.
Second, many ML/DL models operate as black boxes that only predict a label without further explanations, providing little insight into their decision-making processes and hindering effective diagnosis by engineers.
Recent advances leverage large language models (LLMs) for \task~\cite{liu2024logprompt, pan2024raglog}, offering improved interpretability by natural language explanations. However, the massive parameter scale of LLMs intensifies the efficiency problem, which requires expensive hardware and blocks high-volume, real-time inference.

\begin{figure}[t]
    \centering
    \includegraphics[width=0.99\columnwidth]{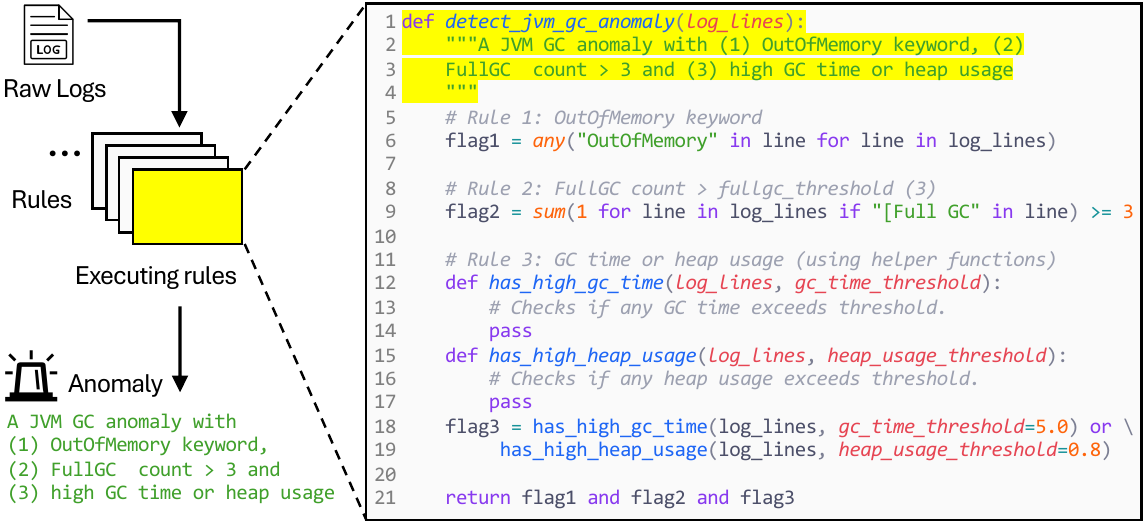}
    \vspace{-6pt}
    \caption{A Python implementation of rule code for a JVM GC anomaly~\cite{zhao2021empirical}.}
    \label{fig:introduction-example}
    \vspace{-15pt}
\end{figure}

Due to these limitations, in real-world practice, many organizations continue to rely on simple rule-based systems for anomaly detection~\cite{ma2025practitioners, zhao2021empirical, he2022empiricalMicrosoft, elastic}, such as keyword matching~\cite{shang2013assisting, yen2013beehive} and timestamp analysis~\cite{rouillard2004lisa, he2022empiricalMicrosoft}. While interpretable and efficient, these systems demand significant manual effort and expert knowledge to develop and maintain, and struggle to scale across diverse and evolving software environments~\cite{zhang2024automl}.

In this paper, we propose a new paradigm for \task by automatically synthesizing executable rules, which can meet the requirements of interpretability and inference efficiency, while preserving high detection accuracy. 
Specifically, each rule is defined as a self-contained Python function that encapsulates detection logic, providing both readability and execution efficiency. Figure~\ref{fig:introduction-example} shows an example rule function to detect the JVM Garbage Collection (GC) anomaly, which contains the appearance of keywords ``OutOfMemory'', the increase in ``FullGC'' template (template count), and the rise of GC time cost and heap space (variable value)~\cite{zhao2021empirical}. The rule function detects this anomaly by returning a \texttt{True} flag for an input log window. Given such a prediction, engineers can instantly alarmed by the triggered rule function and its documentation to understand the anomaly and obtain mitigation insights. 
Additionally, since the rule functions apply simple logic directly to raw log data, they are computationally lightweight and require no expensive GPU hardware while processing streaming online logs.

To synthesize the rules, we propose \nm, an automatic \underline{\textbf{Code}} synthesis framework of rules for log-based \underline{\textbf{A}}nomaly \underline{\textbf{D}}etection powered by LLMs.
\nm operates in two stages: an \textit{offline rule synthesis} phase and an \textit{online monitoring} phase. 
In the \textit{offline} phase, \nm leverages LLMs to analyze labeled log data, recognize anomaly patterns, and automatically generate executable Python rule functions. Specifically, \nm constructs rule functions through three key steps. 
(1) Hierarchical clustering is applied to frequent tokens to group similar log windows. This step organizes large volumes of logs into semantically coherent clusters and reduces the search space for rule generation.
(2) Within each cluster, \nm performs anchor-grounded sampling to form contrastive groups of windows. Specifically, a representative anchor window is selected, and its most similar normal and abnormal windows are retrieved to construct contrastive windows. These windows share similar contents and structures, but differ in discriminative features, allowing LLMs to effectively compare normal and abnormal behaviors and generate meaningful rule logic.
(3) \nm then employs an agentic workflow to synthesize, test, and refine rule functions. The workflow iteratively generates Python code from the contrastive windows, verifies correctness and generalizability, and repairs or improves the code until a reliable rule function is obtained.
In the \textit{online} phase, the synthesized rules are deployed to efficiently classify streaming logs in real time, enabling interpretable and low-cost anomaly detection.
Specifically, streaming logs are grouped into windows and sequentially evaluated by the synthesized functions: windows matched by normal rules are labeled as ``normal'', those matched by abnormal rules are labeled as ``abnormal'', and unmatched windows are conservatively treated as normal to minimize false alarms.

We evaluated the performance of \nm against a range of ML-based, DL-based, LLM-based, and rule-based baselines on three large-scale public datasets for \task, including BGL, Hadoop, and Thunderbird. The results show that \nm outperforms all baselines with limited synthesis cost. For example, when equipped with \textit{gpt-5-mini}, \nm surpasses the current state-of-the-art AdaptiveLog~\cite{ma2025adaptivelog} by 3.6\% F1 while costing only 4 USD per dataset. 
Our efficiency analysis also shows that \nm processes the largest Thunderbird dataset faster than all baselines, facilitating efficient monitoring of large-scale streaming log data in production environments.
Additionally, our comprehensive ablation studies on the method designs, parameter configurations, and LLM backbones confirm the effectiveness of \nm.
Furthermore, our analysis on the characteristics of rule functions show that keyword matching is the most prevalent type of rule, which is consistent with prior findings~\cite{zhao2021empirical}.
Overall, our evaluation shows the potential of \nm for deployment in real-world production systems, where explainability, efficiency, automation, and generalizability are critical concerns.

To sum up, the contributions of this work are as follows:
\begin{itemize}[leftmargin=*]
    \item To the best of our knowledge, \nm is the first automatic rule code synthesis method for \task. We propose encapsulating rules as Python functions and use LLMs to synthesize their code. 
    \item To enable efficient rule synthesizing, we propose hierarchical clustering and anchor-grounded sampling to collect contrastive log windows for LLMs. To avoid overfitting and reduce code errors, we propose an LLM-based agentic workflow to synthesize more generalizable rule functions.
    \item We evaluate \nm on three large-scale datasets. Results show that \nm outperforms state-of-the-art methods by 3.6\% F1 on average while offering superior interpretability and efficiency in terms of inference speed and cost.
\end{itemize}

\vspace{-2mm}
\section{Background and Motivation}

\subsection{Existing Paradigms for Log-based Anomaly Detection}
Log-based anomaly detection (\task) aims to identify abnormal behaviors of software systems by analyzing logs, which is a crucial step in maintaining software reliability~\cite{he2016experience, he2021logsurvey}. Over the years, approaches to \task have evolved into three distinct paradigms, each with a unique set of strengths and weaknesses, as summarized in Table~\ref{tab:background-paradigm_comparison}.

\begin{table*}[h]
\captionsetup{justification=centering}
\centering
\caption{Common Rules Used in Practical LAD. }
\label{tab:background-rules}
\resizebox{0.95\textwidth}{!}{%
\begin{tabular}{l p{1\textwidth}}
\toprule
\textbf{Rule Type} & \textbf{Explanation and Examples} \\
\midrule
Keyword & Detects anomalies via keywords including blacklist (\eg fatal errors), whitelist exceptions, and compliance keywords (\eg sensitive info).
\\

Event Count & Monitor per-event count, coverage of top events, long-tail proportion. Flag significant surge or drop versus seasonal baseline (\eg, >3$x$ or >3$\sigma$). 
\\

New Log Pattern & Identify log events never seen before, \eg new "database timeout" event in APP logs, new "port error" event in device logs. 
\\

Event Sequence & Validate per-trace/request FSM: step order, missing steps, and timeouts; session conversation (Start–End pairing). 
\\

Variables & 
Analyzes numerical variable distributions (\eg response time p95 rises from 100ms to 500ms) and categorical proportion (\eg sudden 10\% increase in HTTP 5xx error proportion).
\\

Threshold & 
Sets thresholds for error rate, spikes/drops, or ratios, \eg >50 ERROR logs per minute, no heartbeat logs for a component in 5 minutes.
\\

Composition & Typically, to enhance performance and adapt to the dynamic environment, engineers tend to spend significant efforts to maintain and update rules constantly (e.g., adding new ''AND'' and ''OR'' conditions).\cite{zhao2021empirical}
\\

\bottomrule
\end{tabular}
}
\vspace{-6pt}
\end{table*}

\noindent \textbf{Machine Learning (ML) and Deep Learning (DL) Based Methods:}
To reduce the manual burden of log analysis, many academic studies have focused on ML-based methods~\cite{ali2025mlLOD}. 
These approaches automatically extract features from raw logs and train classifiers like PCA~\cite{xu2009detecting}, SVM~\cite{liang2007ibmSVM} to distinguish between normal and abnormal behaviors. 
DL-based methods~\cite{le2022DeepLogADStudy} leverage deep neural networks to learn complex features such as word semantics~\cite{guo2021logbert, le2021neurallog} and sequence order~\cite{du2017deeplog,meng2019loganomaly}, which often achieve higher accuracy than ML models.
However, their adoption in industrial settings remains limited for two primary reasons. First, they are notorious for their poor explainability. Neural networks act as ``black boxes'' that output a label without a clear and human-understandable reason~\cite{zhao2021empirical, he2022empiricalMicrosoft}. Second, the inference efficiency can be a bottleneck, as large-scale matrix computations in DL-based methods can be slow and resource-intensive. Additionally, the preprocessing and parsing steps to extract features required in ML-based methods~\cite{khan2023impact, shin2021theoretical} can also be complex and slow, especially when dealing with more complex and high-volume logs consistently produced in online software systems~\cite{he2022empiricalMicrosoft, huang2024lofi} (see also \S\ref{sec:expr-rq4-time} for empirical evidence).

\noindent \textbf{LLM-based Methods:}
More recently, LLMs have been applied in \task; they can read raw log messages and make more accurate and explainable predictions by providing additional justifications for decision making in natural language~\cite{liu2024logprompt, zhang2025aiopssurvey}. 
This can be achieved due to the strong textual understanding and reasoning abilities of LLMs, which have been trained on a massive textual corpus with billions of parameters~\cite{liu2024logprompt, ji2024adapting}. 
Although benefiting from improving effectiveness and interpretability, the billions of parameters in LLMs intensify the efficiency problem, which requires expensive hardware like GPUs and incurs prohibitive costs for real-time, high-volume log processing~\cite{ma2025adaptivelog}. 
Even though some companies offer payable API services to reduce the need of local deployment, the cost will be unaffordable in the long run. For example, it is estimated that processing an hour of 200M lines of logs (10 tokens per log) using GPT-4 could cost 60000 dollars~\cite{ma2025adaptivelog}.
Furthermore, while automated, LLMs can struggle with generalizability, often generating detection logic that overfits to specific details (\eg a specific IP address) in the provided examples rather than capturing broader, more robust anomaly patterns~\cite{huang2025logrules}.

\begin{table}[t]
\centering
\caption{Comparison of Log Anomaly Detection Paradigms}
\label{tab:background-paradigm_comparison}
    \resizebox{0.485\textwidth}{!}{
\begin{tabular}{|l|cccc|}
\hline
\textbf{Paradigm} & \textbf{Explainability} & \textbf{Efficiency} & \textbf{Automation} & \textbf{Generalizability} \\
\hline
ML/DL-based & $\times$ & $\times$ & $\checkmark$ & $\times$ \\
LLM-based & $\checkmark$ & $\times$ & $\checkmark$ & $\times$ \\
Rule-based (Manual) & $\checkmark$ & $\checkmark$ & $\times$ & $\times$ \\
\hline
\textbf{\nm (Ours)} & $\checkmark$ & $\checkmark$ & $\checkmark$ & $\checkmark$ \\
\hline
\end{tabular}
}
\end{table}

\noindent \textbf{Rule-based Systems:}
In practice, many large cloud companies still rely on simple, manually-crafted rule-based systems for \task~\cite{he2022empiricalMicrosoft, zhao2021empirical}. As shown in a recent survey among \task practitioners~\cite{ma2025practitioners}, 75.5\% of the tools they used in practice are still based on heuristic rules, \eg keyword matching. 
These rules are often initially created based on engineers' operational experience, where they summarized some commonly-appeared  anomaly patterns in logs during their operation practice (\eg ``OutOfMemory'', ``timeout''), and then continuously updated during online deployment by analyzing identifying emerging patterns from incident (\eg adding new ``AND'' and ``OR'' conditions)~\cite{zhao2021empirical, liu2023sealog}. 
Table~\ref{tab:background-rules} shows examples of rules used in practice~\cite{he2022empiricalMicrosoft, zhao2021empirical}. 
The enduring popularity of this method in practice stems from its two strengths: 
(1) high explainability where engineers can easily understand the rules due to transparent and straightforward patterns (\eg the occurrence of some events)~\cite{zhao2021empirical, he2022empiricalMicrosoft}; and 
(2) excellent inference efficiency where the rules contain simple anomaly logics (\eg if-else) and can be executed without using expensive hardware~\cite{ma2025practitioners, liu2023sealog, elastic, loki}. 
However, this paradigm's critical weakness is its complete lack of automation and poor generalizability. 
Due to the complex business logic and specific system designs, manually creating and tuning rules is a time-consuming, error-prone process that relies heavily on expert experience. Additionally, 
such rules are often tailored to a specific component, making them hard to scale and adapt to modern software systems that become more complex and distributed.

\subsection{Motivation: A New Paradigm of Automated Rule Synthesis}
The current state of \task reveals a significant gap between academic research and industrial needs. ML, DL and LLM-based methods offer automation but fail to meet the practical requirements for efficiency and interpretability that system engineers demand. In contrast, the rule-based systems used in industry are efficient and interpretable but are not automated and do not generalize well.

This work aims to bridge this gap by creating a solution that achieves explainability, inference efficiency, and automation at one time. We propose a new paradigm that combines the advantages of existing approaches by repurposing the power of LLMs. Rather than using them directly for anomaly detection, our core idea is to leverage their advanced code generation and pattern-inference capabilities to automatically synthesize high-quality and executable detection rules. Recent studies have shown the effectiveness of LLMs in recognizing log patterns~\cite{huang2025lunar} and generating code~\cite{hou2024llm4sesurvey} by following specified instructions. Thus, we believe that LLMs also have the potential to generate accurate and interpretable rules for \task. 
Figure~\ref{fig:introduction-example} shows an example of code rules in Python. By doing so, we aim to deliver a system with the following advantages:
\begin{itemize}[leftmargin=*]
    \item \textbf{Explainable}: Since synthesized rules correspond to human-understandable code, they offer transparent detection logic as manually-written rules.
    \item  \textbf{Efficient}: Once generated, the code rules are lightweight and can be executed for online detection without the computational overhead of LLMs.
    \item \textbf{Semi-automated}: LLMs are used to automatically infer rules from labelled log windows, eliminating the need for manual rule creation. 
\end{itemize}

Ultimately, this approach harnesses the power of LLMs to automate the creation of practical, efficient, and interpretable rules, providing a novel solution that aligns with the real-world demands of software reliability engineering.

\subsection{Challenges of Applying LLMs for \task Rule Synthesis}

However, applying LLMs for log-based anomaly detection rule generation has the following challenges:

\begin{itemize}[leftmargin=*]
    \item \textbf{Challenge 1: Effective and efficient sampling of windows:} 
    Due to the sheer volume of log windows, we cannot feed all windows to LLM inputs~\cite{he2021logsurvey}. Consequently, a group of windows can be sampled. However, constructing effective window groups is challenging due to the difficulty in balancing similarity and variability within groups. The LLMs can fail to extract common patterns and generate a rule code that overfit to each pattern if provided with windows sharing poor similarity. 
    Additionally, the sheer volume of windows means that we need an efficient methods to scale, which further complicates the sampling.
    \item \textbf{Challenge 2: Complexity and diversity of rules:} A software system can have a large number of complex patterns for both anomaly (\eg a component fails) and normality (\eg normal execution of events)~\cite{zhao2021empirical}. 
    The diversity and complex inter-relationships of rules can complicate the LLM-based construction and impede the readability. 
    For example, simply composing all rule patterns in one function can increase the chances of generation errors. 
    Thus, appropriation organization of a set of rules is important. 
    
    \item \textbf{Challenge 3: Generalizability and correctness of rules:} Although provided with multiple log windows, the LLMs can still generate rule code that matches specific details within the logs (\eg matching a specific hex parameter or ip address), impeding the generalizability of the rules. Additionally, LLMs may still generate rule code that contains syntax errors, despite being pretrained on billions of lines of source code~\cite{chen2021evaluating}. 
\end{itemize}

\vspace{-6pt}
\section{Methodology}
To address the challenges, we introduce \nm for \task rule synthesis. Figure~\ref{fig:method-framework} shows the overall framework of \nm, comprising two phases: (1) an \textit{offline synthesis phase}, where \nm synthesizes executable detection rules using LLMs and labeled training log windows (\S\ref{sec:method-clustering}--\ref{sec:method-validation}), and (2) an \textit{online detection phase}, where the synthesized rules are applied to streaming logs for real-time detection (\S\ref{sec:method-online}).

\vspace{-6pt}
\subsection{Overview of Offline Synthesis}\label{sec:method-overview}
\begin{figure*}[t]
    \centering
    \includegraphics[width=0.95\textwidth]{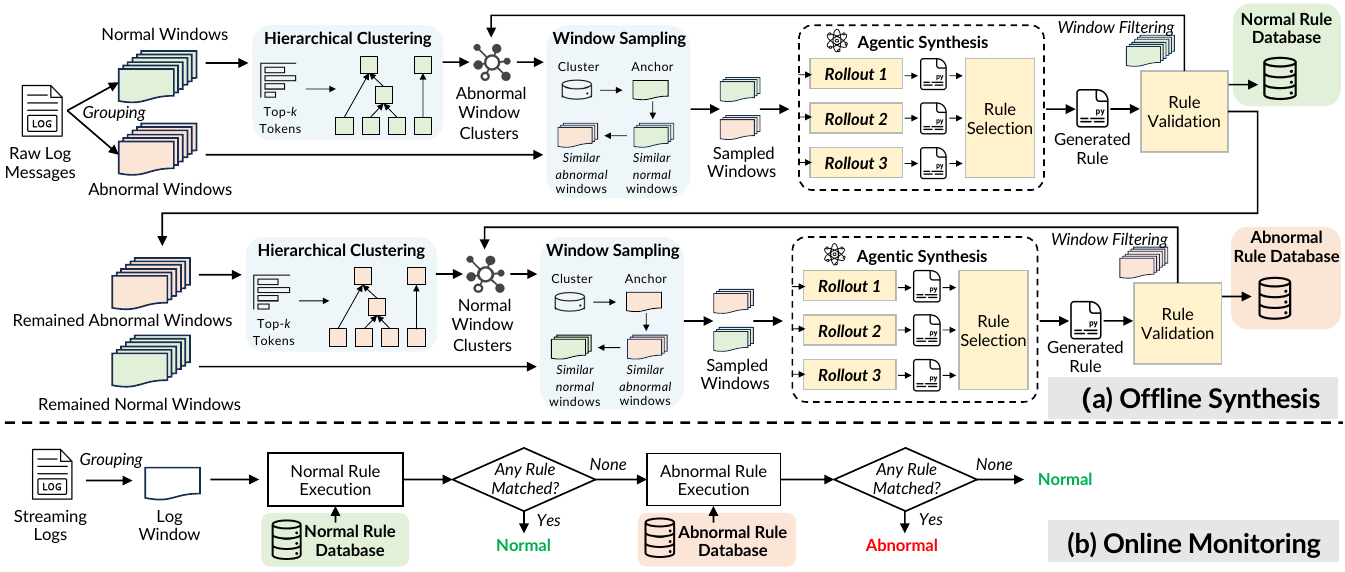}
    \vspace{-5pt}
    \caption{The overall workflow of \nm.}
    \label{fig:method-framework}
    \vspace{-15pt}
\end{figure*}

Figure~\ref{fig:method-framework}-(a) shows the overall synthesis framework.
To address the complexity and diversity of anomaly patterns, \nm uses a dual-rule synthesis strategy, generating two separate sets of rules in sequential orders: \textit{normal rules} and \textit{abnormal rules}. Normal rules capture recurrent patterns in normal log windows, while abnormal rules target anomalous patterns. This separation of rule sets disentangles the underlying pattern spaces, which reduces rule complexity, improves interpretability, and simplifies rule ordering. 
To address the challenge of effective and efficient window sampling, we use hierarchical agglomerative clustering (HAC) and propose a anchor-grounded window sampling method. The HAC efficiently groups similar log windows based on the top-$k$ most frequent tokens and reduces computational overhead(\S\ref{sec:method-clustering}). The anchor-grounded sampling method selects a representative anchor window and retrieves similar windows to form contrastive window groups for LLM inference (\S\ref{sec:method-sampling}).
To address the generalization challenge, we introduce an agentic synthesis workflow inspired by group relation optimization~\cite{shao2024deepseekmath} (\S\ref{sec:method-agent}). This workflow leverages multiple LLM-based agents in a rollout framework to generate, test, repair, and refine candidate rule code for each sampled group. Each candidate rule is first locally tested for syntax and execution correctness. If errors are detected, a repair agent iteratively fixes the code. To mitigate overfitting to specific log details, a generalizability validation module evaluates each rule on a broader set of windows. Rules that fail to meet a predefined generalizability threshold are sent to a refine agent, which prompts the LLM to revise the code for improved generalization, such as by abstracting specific parameters or simplifying conditions. This iterative, multi-agent process continues until a rule passes all quality checks and is deemed robust.
Once validated, the synthesized rules are stored in a rule database, and windows detected by these rules are filtered out of clusters to avoid redundant synthesis (\S\ref{sec:method-validation}). The remaining clusters move to the next epoch, which involves a full cycle of window sampling and agentic synthesis. The same workflow is symmetrically applied to both normal and abnormal rule sets, with tailored parameters and prompts for each. This systematic approach ensures that the final rules are robust, interpretable, and effective for anomaly detection across diverse log environments.

\subsection{Hierarchical Window Clustering}\label{sec:method-clustering}

Directly sampling similar log windows from the entire dataset is computationally prohibitive due to the massive volume of log data, which often contains millions of log lines. To address this challenge, \nm first partitions similar log windows into clusters, where each cluster contains windows sharing similar log tokens. This clustering step ensures that dissimilar windows, \ie unlikely to share the same anomaly or regularity patterns, are separated, thereby reducing redundant computation and improving the efficiency of subsequent rule synthesis. To achieve efficient and scalable clustering, we propose a hierarchical agglomerative clustering (HAC) method based on window-level token features. Specifically, \nm first obtains top-$k$ frequent tokens of each window to form initial clusters and then iteratively merge the clusters in a bottom-up way. 
In the following, we first introduce the method for clustering normal windows and then introduce the mirrored method for abnormal windows.

\subsubsection{Clustering Normal Windows} Before clustering, we first extract the most frequent tokens in each window as its representations. Specifically, each log message within a window is tokenized by whitespace, and the frequency of each token is counted. We then select the top-$k_{\mathit{line}}$ most frequent tokens from each log message to represent each log. This step filters out less common tokens to reduce noise, which is a common practice in log analysis~\cite{huang2025lunar, he2017drain}. Next, for each window, we aggregate the top-$k_{\mathit{line}}$ tokens across all contained log messages and select the $k_{\mathit{window}}$ most frequent tokens as the window's feature. To adapt to varying window sizes, $k_{\mathit{window}}$ is dynamically set as $k_{\mathit{window}} = \mathtt{int}(\alpha \cdot L)$, where $L$ is the average number of logs per window and $\alpha$ is a scaling factor.

Based on these window-level token features, we perform hierarchical agglomerative clustering in a bottom-up manner. Initially, windows with identical top-$k_{\mathit{window}}$ tokens are grouped into singleton clusters. Then the clustering process proceeds iteratively: at each iteration $r$ ($r=1,2,\ldots,m$), we randomly sample pairs of clusters and compute their similarity based on the overlap of their feature tokens. If two clusters share at least $k_{\mathit{window}} - r$ tokens, they are merged into a higher-level cluster, with the intersection of their token sets serving as the new cluster feature. This iterative merging continues until no further clusters can be merged or a predefined maximum number of iterations is reached. This approach gradually relaxes the similarity criterion, allowing clusters to grow while maintaining semantic coherence.

\subsubsection{Clustering Abnormal Windows} To synthesize abnormal rules, we apply the same hierarchical clustering method, but target abnormal windows instead of normal ones. The clustering parameters and merging logic remain consistent, ensuring symmetry between the two rule synthesis workflows.

This hierarchical clustering strategy effectively organizes log windows into semantically coherent groups, enabling efficient and targeted sampling for subsequent rule synthesis.

\subsection{Anchor-grounded Window Sampling}\label{sec:method-sampling}
After clustering, the next step is to sample representative groups of log windows for LLM-based comparison and rule inference, which are termed contrastive windows. An effective sampling strategy should select windows that are both diverse and similar, ensuring that both normal and abnormal windows share sufficient intra-group and inter-group similarity. This enables the LLM to identify nuanced differences between normal and abnormal patterns, thereby synthesizing generalizable detection rules. To achieve this, we propose an anchor-grounded sampling method, which selects a representative anchor window and samples similar windows from both the same and opposite categories.
In the following, we first describe the sampling process for normal rule synthesis, followed by the mirrored procedure for abnormal rules.

\subsubsection{Contrastive Windows with Normal Anchor Window}
In each epoch, we begin by selecting the largest cluster of normal windows. Within this cluster, we identify an anchor window that exhibits the highest diversity in log messages, as measured by the diversity score $\mathtt{diversity} = \frac{\#\,\mathtt{unique\,tokens}}{\#\,\mathtt{total\,tokens}}$. A highly diverse anchor window is more likely to represent the overall distribution of the cluster and helps mitigate overfitting to narrow patterns, promoting the synthesis of rules that generalize to long-tail cases.

After selecting the anchor window, we sample $w-1$ additional normal windows from the same cluster based on their similarity to the anchor. We use the Jaccard similarity metric, defined as $\mathit{JS}(w_1, w_2) = \frac{|\mathit{Topk}(w_1) \cap \mathit{Topk}(w_2)|}{|\mathit{Topk}(w_1) \cup \mathit{Topk}(w_2)|}$, where $\mathit{Topk}(w)$ denotes the set of top-$k_{\mathit{window}}$ tokens for window $w$. We select the $w-1$ windows with the highest similarity scores above a threshold $\theta^{\mathit{anc}}_{\mathit{nor}}$, together with the anchor, to form a set of $w$ candidate normal windows. Selecting windows with high similarity to the anchor ensures that the sampled group captures the major patterns in the cluster, while avoiding windows with low similarity, which are less likely to share the same normal patterns and may represent rare or unrelated system behaviors. This approach helps the LLM focus on learning the essential characteristics of normal behavior, improving the generalizability of the synthesized rules.

To facilitate contrastive learning, we then sample $w$ abnormal windows that are most similar to the selected normal windows. For each abnormal window, we compute its average Jaccard similarity to the $w$ normal windows and select those with the highest scores.
This ensures that the sampled abnormal windows share sufficient overlap with the normal windows, enabling the LLM to effectively discriminate between normal and anomalous patterns. Notably, since abnormal windows are typically much fewer than normal windows, the computational overhead of this enumeration is acceptable in practice.

Finally, a contrastive set of $w$ normal windows and $w$ abnormal windows are obtained and used for following synthesis.

\subsubsection{Contrastive Windows with Abnormal Anchor Window}
For abnormal rule synthesis, we apply a mirrored anchor-grounded sampling strategy. Specifically, we select the largest cluster of abnormal windows and identify the anchor abnormal window with the highest diversity score. We then sample $w-1$ similar abnormal windows from the same cluster using the Jaccard similarity. Finally, we sample $w$ normal windows from all deduplicated normal windows that are most similar to the selected abnormal windows, using the same similarity thresholding approach.

Anchor-grounded sampling ensures that each sampled group contains representative and diverse windows from both categories, enabling synthsizing accurate and generalizable rules.

\subsection{Agentic Synthesis Workflow}\label{sec:method-agent}
Upon obtaining contrastive windows, \nm utilizes large language models (LLMs) to synthesize detection rule code. Effective rule code should be interpretable by engineers, generatable by LLMs, and executable during detection. Thus, we target Python as the rule language, which has excellent readability and frequent usage in LLM-based code generation tasks~\cite{zhang2023surveyllm4se}, though our approach can be extended to other languages by modifying prompt instructions.

As LLMs are not specifically tuned for \task rule synthesis, they may generate suboptimal rules with syntax errors or overfitting to specific log details (\eg matching a specific hex value or IP address). 
To address these challenges, we adopt an agentic synthesis workflow comprising multiple rollouts and specialized agents for generation, repair, and refinement. This multi-agent approach leverages the inherent stochasticity of LLMs to generate diverse candidate rules, increasing the likelihood of obtaining robust and generalizable solutions.

\begin{figure}[t]
    \centering
    \includegraphics[width=0.99\columnwidth]{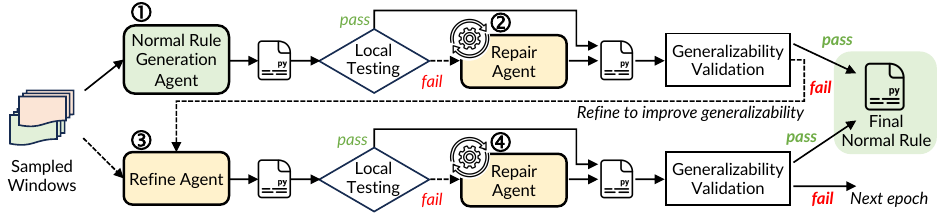}
    \caption{The rollout process of \nm for normal rules synthesis. }
    \label{fig:method-llm-agent}
    \vspace{-10pt}
\end{figure}

\subsubsection{Agentic Rollout}\label{sec:method-agent-rollout}
Inspired by group relation optimization~\cite{shao2024deepseekmath}, we perform $r$ independent rollouts, each generating a candidate rule function. Figure~\ref{fig:method-llm-agent} shows the agentic framework for normal rule synthesis, comprising three LLM-based agents and two validation modules for syntax and generalizability.

\paragraph{Synthesizing Normal Rules} The agentic synthesis process to rollout one normal rule is as follows:

\noindent \textbf{Normal Rule Generation Agent.}
The agent generates a complete Python function (\ie \texttt{detect\_regularity()}) that takes a list of log message strings as input and returns \textsc{TRUE} if the window is normal, and \textsc{FALSE} otherwise. To guide LLMs to generate detection code in the expected format, we design a comprehensive prompt including detailed task instructions, output constraints, and contrastive normal and abnormal windows. 
Instructions provide guidance on discriminating rules through comparison, which guide LLMs to first generate a docstring describing the rule in natural language and then implement the function following the docstring. Output constraints ensure the generated code is well-documented, correctly formatted, and easily extractable.

\noindent \textbf{Local Testing.}
Despite guided by clear instructions in the prompt, LLMs can still produce anomaly detection rules  with syntax errors or inaccuracies. To verify syntax and execution correctness of code, we perform local testing using unit tests~\cite{zhu1997softwareunittest} on local test cases, \ie $2w$ sampled contrastive windows. If the rule function produces the expected Boolean anomaly labels for all $2w$ windows without errors, it \textit{passes} local testing and proceeds to generalizability validation. Otherwise, it is sent to the repair agent for correction.

\noindent \textbf{Repair Agent.}
Despite performing well in various Python code generation tasks~\cite{chen2021evaluating, austin2021program}, LLMs may still generate rule code with syntax errors. To fix the errors, we introduce a repair agent.
Specifically, we prompt the LLM with the instructions, faulty function, error messages, and misclassified windows. The agent iteratively generates corrected versions until all local testing errors are resolved.

\noindent \textbf{Generalizability Validation.}
Since LLMs are not specifically optimized for \task rule synthesis, they often generate rules focusing on specific details within logs (\eg matching a specific hex values or IP address), which impairs rules' generalizability. To address this issue, we propose a validation module to assess their generalizability on a broader set of windows. 
Specifically, we apply unit tests to all remaining normal windows in the training set and compute the proportion of correctly identified normal windows. If this proportion exceeds a predefined threshold $g_{\mathit{nor}}$, the rule is deemed generalizable and accepted as the final normal rule for the epoch. Otherwise, it is sent to the refine agent for further improvement.

\noindent \textbf{Refine Agent.} 
The refine agent improves rules' generalizability by generating a refined rule function. Specifically, we design a comprehensive refinement prompt containing detailed task instructions, output constraints, the current code and contrastive windows. 
The refinement instruction explains the cause of poor generalizability (\eg overfitting to specific parameters that only appear in sampled windows) and requests more generalized logic in the generations (\eg using regular expressions or simplifying conditions).
Once the refined code is generated, it undergoes the same local testing and generalizability validation as before. If it passes, the rule is stored; otherwise, the epoch is discarded and a new round of window sampling and synthesis is initiated.

\paragraph{Synthesizing Abnormal Rules} 
For abnormal rule synthesis, we employ a mirrored agentic workflow to that of normal rules. The process consists of the abnormal rule generation agent producing candidate functions, followed by local testing, repair, and generalizability validation. This symmetry ensures that both normal and abnormal detection rules are robust, generalizable, and suitable for deployment.
 
\noindent \textbf{Abnormal Rule Generation Agent.}
The abnormal rule generation agent produces a Python function (\ie \texttt{detect\_anomaly()}), which takes a list of log message strings as input and returns \textsc{TRUE} if the window is abnormal, and \textsc{FALSE} otherwise. To prompt LLMs, we design a comprehensive prompt including task instructions, output constraints and contrastive windows. 
The task instructions provide guidance on identifying anomalous patterns, with examples such as those in Table~\ref{tab:background-rules},  and require the LLM to first generate a descriptive docstring before implementing the function. Output constraints ensure the generated code is well-documented, correctly formatted, and easily extractable. 
This modular design allows the abnormal rule generation agent to focus on the unique aspects of anomaly detection, ensuring that the synthesized rules are both accurate and interpretable. 

\noindent \textbf{Abnormal Rule Workflow.}
After generation, abnormal rule functions undergo local testing on contrastive windows, followed by iterative repair if necessary. Then generalizability validation is performed on remaining abnormal windows in the training set, with a threshold $g_{\mathit{abn}}$ determining rule acceptance. If validation fails, the rule is sent to the refine agent for enhancement, mirroring the process used for normal rules.

\subsubsection{Rule Selection}\label{sec:method-agent-rule-selection}
After obtaining $r$ candidate rule functions from rollouts, we select the best one for this epoch using a two-stage rigorous selection process. 
First, we apply unit tests to all candidates on the validation set and eliminate rules that misclassify any window of the opposite label. For example, a normal rule that misclassifies at least one abnormal window as normal will be removed from candidates. 
Next, from remaining candidates, we select the rule with the highest accuracy in identifying windows pertaining to the target category. 
This method ensures that the selected rule maximizes coverage of normal (or abnormal) windows while minimizing false alarms. It's worth noting that the selection module can be extended to more criteria, \eg inference time or resource usage.

\vspace{-8pt}
\subsection{Rule Application and Epoch Management}\label{sec:method-validation}
After obtaining a validated rule function by agentic synthesis, we apply the rule to the training windows to identify and filter out those that are successfully detected. This process prepares the dataset for the next synthesis epoch by excluding windows that have already been covered by existing rules.

\textit{Normal Rules.}
In the normal rule synthesis workflow, windows classified as ``normal'' by the newly synthesized rule are excluded from the current clusters, indicating successful coverage of these patterns. The validated rule is then added to the normal rule database, which maintains a list of all synthesized normal rule functions. The synthesis process continues until either the number of remaining normal windows falls below a threshold $R_{\mathit{nor}}$, or the normal rule database reaches its maximum capacity $D_{\mathit{nor}}$. At this point, normal rule synthesis is terminated, and the remaining normal windows—those not identified by any normal rule—are reserved for subsequent abnormal rule synthesis.

\textit{Abnormal Rules.}
Similarly, in the abnormal rule synthesis workflow, windows classified as ``abnormal'' are removed from the clusters. The process is guided by analogous stopping criteria: synthesis halts when the proportion of remaining abnormal windows drops below $R_{\mathit{abn}}$ or the abnormal rule database reaches its maximum size $D_{\mathit{abn}}$.

This iterative filtering and rule accumulation process ensures that each epoch incrementally covers the pattern space, progressively reducing the number of uncovered windows and avoiding redundant rule synthesis.

\vspace{-8pt}
\subsection{Online Monitoring Phase}\label{sec:method-online}
The synthesized rule databases are seamlessly integrated into the online system monitoring pipeline, enabling efficient and accurate log anomaly detection at scale. After the offline synthesis phase, both the normal and abnormal rule databases are deployed for real-time detection. Figure~\ref{fig:method-framework}-(b) shows the online monitoring framework of \nm.

When new logs arrive, \nm first groups them into log windows using the same strategy as in the synthesis phase. For each incoming window, the system sequentially executes the normal and abnormal rule functions for classification. Specifically, the window is first evaluated by all functions in the normal rule database; if any \texttt{detect\_regularity()} function returns \textsc{TRUE}, the window is labeled as ``normal''. Windows not matched by any normal rule are then evaluated by the abnormal rule database; if any \texttt{detect\_anomaly()} function returns \textsc{TRUE}, the window is labeled as ``abnormal''. Windows that are not classified by either rule set are considered outliers. Given the empirical rarity of anomalies in large-scale log data, we conservatively label such windows as ``normal'' to minimize false positives. This two-stage rule application strategy ensures high precision in anomaly detection while maintaining computational efficiency.

\vspace{-8pt}
\section{Experiment Setup}
We investigate the following research questions (RQs) to  evaluate the effectiveness and efficiency of \nm in \task. These questions ensure a thorough understanding of \nm’s capabilities, limitations, and practical value for industrial anomaly detection. 
\begin{itemize}[leftmargin=*, topsep=0pt]
    \item \textbf{RQ1:} How effective is \nm in \task?
    \item \textbf{RQ2:} How do different clustering and sampling components affect \nm?
    \item \textbf{RQ3:} How do different agentic components affect \nm?
    \item \textbf{RQ4:} How efficient and cost-effective is \nm in monitoring large-scale log data?
    \item \textbf{RQ5:} How can \nm benefit engineers?
\end{itemize}

\vspace{-10pt}
\subsection{Datasets}
We conducted comprehensive experiments on three widely used log-based anomaly detection public datasets: BGL, Thunderbird \cite{oliner2007supercomputers}, and Hadoop \cite{he2020loghub}. The BGL dataset is derived from the operational logs of the Blue Gene/L supercomputing system, which contains 128K processors. The Thunderbird dataset contains over 200 million log messages, collected a supercomputer with 9,024 processors and 27,072 GB of memory. The Hadoop dataset was gathered from a 46-core Hadoop distributed file system spanning five machines.
Considering the huge scale of the Thunderbird dataset, we followed the settings of the previous study \cite{le2021neurallog, le2022DeepLogADStudy, he2025weakly} and selected the earliest 10 million log messages for experimentation. 
A recent study~\cite{landauer2024critical} recommended HDFS~\cite{xu2009detecting} and ADFA-LD~\cite{creech2013generation} datasets for evaluation due to their increased complexity of log anomalies. However, we excluded HDFS due to its focus on session-level anomalies and ADFA-LD as only a preprocessed version of numeric identifiers are available.

\vspace{-10pt}
\subsection{Baselines}
To better evaluate the effectiveness of \nm, we compared it with \textbf{8} open-source state-of-the-art log-based anomaly detectors, including rule-based detectors, machine-learning-based detectors, deep-learning-based detectors, and LLM-based detectors. 
\textbf{3-Sigma} \cite{he2022empiricalMicrosoft} and \textbf{BDA Assist} \cite{shang2013assisting} are extensively employed rule-based methods for anomaly detection, which rely on SRE-predefined rules to identify system anomalies. 
\textbf{LR} \cite{he2016experience}, \textbf{SVM} \cite{he2016experience}, and \textbf{SemPCA} \cite{yang2024try} represent state-of-the-art machine learning-based detectors that employ classification or clustering paradigms to identify anomalies in system execution flows. 
\textbf{LogRobust} \cite{zhang2019Logrobust} and \textbf{NeuralLog} \cite{le2021neurallog} are the leading deep-learning-based methods that utilize supervised learning, employing neural networks to map log sequences to vectors and then using classification-based approaches for anomaly detection. 
\textbf{AdaptiveLog} \cite{ma2025adaptivelog} stands as a cutting-edge LLM-based detector that realizes effective anomaly detection via the synergistic interaction between small and large models.

\vspace{-10pt}
\subsection{Evaluation Metrics}

Following previous work~\cite{zhu2019tools,khan2022guidelines,jiang2023large}, we use precision, recall, and F1-score as evaluation metrics to comprehensively assess the effectiveness of \nm in log-based anomaly detection. Their definitions are as follows: $\mathit{Precision}=\frac{\mathit{TP}}{\mathit{TP}+\mathit{FP}} $, $\mathit{Recall}=\frac{\mathit{TP}}{\mathit{TP}+\mathit{FN}} $, $F1=\frac{2*\mathit{Prec}*\mathit{Rec}}{\mathit{Prec}+\mathit{Rec}}$.

\vspace{-6pt}
\subsection{Environment and Implementation}\label{sec:implementation}
We conducted all the experiments on a Ubuntu 20.04.4 LTS server with 256GB RAM and an NVIDIA A100 40GB GPU. 
The default LLM used in \nm is the latest \textit{GPT-5-mini} model due to its strong performance in coding and instruction following, We invoked the official API provided by OpenAI~\cite{openai-api}. To minimize the randomness introduced by token sampling, we repeated each experiment three times and reported the average performance. Following the setup of existing work~\cite{meng2019loganomaly, le2022DeepLogADStudy, he2025weakly}, we grouped the dataset using a sliding window of size 20, parsed the log messages with Drain, and split the dataset into training, validation, and testing sets in a ratio of 6:1:3. The reason why considering group-wise logs for \task rather than a single log message is due to contextual time dependency existing in logs~\cite{zhao2021empirical}. For the baseline methods used for comparison in the experiments, we utilized their publicly available implementations and hyperparameter settings.

To ensure fair and reproducible evaluation, we adopted consistent parameter settings for both normal and abnormal rule synthesis in \nm. Unless otherwise specified, the number of sampled windows per group ($w$) was set to 5 for both normal and abnormal rules. For feature extraction, we selected the top $k_{\mathit{line}}=2$ tokens from each log line, and used a scaling factor $\alpha=0.5$ to determine the number of top tokens per window. The HAC process was performed with a maximum of $m=4$ merge iterations. During window sampling, the similarity threshold for anchor selection ($\theta^{\mathit{anc}}_{\mathit{nor}}$ and $\theta^{\mathit{anc}}_{\mathit{abn}}$) wass set to 0.2 for both rule types. The rule databases were capped at $D_{\mathit{nor}}=200$ and $D_{\mathit{abn}}=200$ rules for normal and abnormal detection, respectively. Rule synthesis for each epoch terminated when the proportion of remaining windows fell below $R_{\mathit{nor}}=99\%$ for normal rules and $R_{\mathit{abn}}=99.5\%$ for abnormal rules. For agentic synthesis, we used a rollout number of 2 for both normal and abnormal rule generation. We selected all parameter values based on empirical validation; we further analyze their impact in our ablation experiments (\S\ref{sec:expr-rq2}).

\vspace{-6pt}
\section{Evaluation Results}

\begin{table*}[t]
  \centering
  \caption{Accuracy of \nm compared to state-of-the-art baselines (\%)}
\vspace{-6pt}
\resizebox{0.75\textwidth}{!}{%
    \begin{tabular}{ccccccccccc}
    \toprule
    \multirow{2}[4]{*}{\textbf{Paradigm}} & \multirow{2}[4]{*}{\textbf{Model}} & \multicolumn{3}{c}{\textbf{BGL}} & \multicolumn{3}{c}{\textbf{Hadoop}} & \multicolumn{3}{c}{\textbf{Thunderbird}} \\
\cmidrule{3-11}          &       & \textbf{Precision} & \textbf{Recall} & \textbf{F1} & \textbf{Precision} & \textbf{Recall} & \textbf{F1} & \textbf{Precision} & \textbf{Recall} & \textbf{F1} \\
    \midrule
    \multirow{2}[2]{*}{Rule-based} & 3-Sigma & 38.8  & \textbf{99.9}  & 55.9  & 44.6  & \textbf{100} & 61.7  & 19.6  & 95.1  & 32.6 \\
          & BDA Assist & 44.2  & 99.7  & 61.2  & 9.8   & \textbf{100} & 17.8  & 5.3   & \textbf{99.8}  & 10.1 \\
    \midrule
    \multirow{3}[2]{*}{ML-based} & LR    & 88.9  & 91.9  & 90.4  & 58.8  & 94.7  & 72.5  & 94.8  & 18.0  & 30.2 \\
          & SVM   & 55.0  & 90.5  & 68.4  & 70.0  & 90.6  & 79.0  & 70.0  & 81.6  & 75.4 \\
          & SemPCA & 38.8  & 65.1  & 48.6  & 30.5  & 75.0  & 43.4  & 24.2  & 60.9  & 34.7 \\
    \midrule
    \multirow{2}[2]{*}{DL-based} & NeuralLog & 85.7  & 88.2  & 87.0  & 87.2  & 89.8  & 88.5  & 65.3  & 43.9  & 52.5 \\
          & LogRobust & 89.7  & 91.7  & 90.7  & 45.6  & 94.4  & 61.5  & 65.4  & 81.7  & 72.6 \\
    \midrule
    LLM-based & AdaptiveLog & 92.3  & 92.7  & 92.5  & 92.2  & 91.7  & 91.9  & 81.9  & 81.7  & 81.8 \\
    \midrule
    
    & \textbf{\nm} (\textit{gpt-5-mini}) & 96.2  & 89.4  & 92.6  & \textbf{100} & 94.7  & \textbf{97.3} & \textbf{99.3} & 77.5  & \textbf{87.0} \\
    & \textbf{\nm} (\textit{gpt-5}) & \textbf{99.0}  & 90.9  & \textbf{94.8}  & \textbf{100} & 93.2  & 96.5  & 96.9  & 71.2  & 82.1 \\
    
    \bottomrule
    \end{tabular}%
}
  \label{tab:expr-main-table}%
\vspace{-10pt}
\end{table*}%

\subsection{RQ1: How effective is \nm in \task?}\label{sec:expr-rq1}
\noindent \textbf{Setup:} 
To address this RQ, we evaluate the effectiveness of \nm in \task by benchmarking its accuracy against eight state-of-the-art detectors spanning rule-based, ML-based, DL-based, and LLM-based approaches. 
Table~\ref{tab:expr-main-table} reports precision, recall, and F1 scores for all methods, with the best results highlighted in \textbf{bold}.

\noindent \textbf{Results:} 
Overall, \nm demonstrates consistently superior F1 scores across all datasets, demonstrating its robust and generalizable performance in \task. Specifically, \nm with \textit{gpt-5} achieves the highest F1 score on BGL (94.8\%), while \nm with \textit{gpt-5-mini} achieves the best F1 scores on Hadoop (97.3\%) and Thunderbird (87.0\%). Notably, both variants of \nm deliver exceptionally high precision (up to 99.0\% on BGL, 100\% on Hadoop, and 99.3\% on Thunderbird), indicating its strong capability to minimize false positives and enhance operational reliability.

Compared to traditional rule-based methods (\ie 3-Sigma and BDA Assist), \nm achieves substantially higher precision and F1 scores. While rule-based approaches yield near-perfect recall ($\approx$ 100\%), they suffer from low precision due to excessive false alarms. In contrast, \nm detects anomaly more conservatively, resulting in fewer false alerts and reduced alert fatigue—a critical consideration in practical deployment. The complementary strengths of rule-based methods (high recall) and \nm (high precision and F1) suggest that hybrid approaches may further improve anomaly detection performance, which we identify as a promising direction for future work.

Among all baselines, the LLM-based AdaptiveLog achieves competitive F1 scores, benefiting from advanced language models that better capture log semantics. However, AdaptiveLog incurs significant computational overhead, as each log window must be processed by a large language model. In contrast, \nm leverages synthesized Python code rules for online detection, enabling efficient detection and higher accuracy across all datasets (\eg $+0.1\%$, $+5.4\%$, and $+5.2\%$ F1 improvements on BGL, Hadoop, and Thunderbird, respectively, compared to AdaptiveLog). These results highlight the advantages of \nm in delivering accurate, interpretable, and computationally efficient anomaly detection solutions suitable for real-world industrial deployment.

\summary{\textbf{Answer to RQ1}: \nm consistently outperforms baselines in \task, achieving the highest F1 and precision across all evaluated datasets, indicating its strong potential for reliable and accurate deployment in real-world systems. }

\vspace{-6pt}
\subsection{RQ2: How do different clustering and sampling components affect \nm?}\label{sec:expr-rq2}
\subsubsection{Design Analysis on Components}\label{sec:expr-rq2-design}
\nm uses hierarchical clustering and anchor-grounded sampling to collect contrastive windows for LLMs to infer rules. In this RQ, we first assess the individual contribution of \nm's designs in clustering and sampling. To achieve this, we conduct an ablation study by systematically modifying or removing key components. 
Specifically, we evaluate: (1) clustering, (2) sampling, and (3) anchor selection strategy.
For \textit{clustering}, we compare our hierarchical window clustering with a simplified variant that clusters windows based on identical top-$k$ tokens, where $k$ is determined by the window size and the number of merging iterations.
For \textit{sampling}, we assess three alternatives: (a) replacing anchor-grounded sampling with random sampling of $2w$ windows, (b) half-random sampling (randomly selecting windows of opposite labels), and (c) removing contrastive sampling (sampling only anchor and its similar windows in the same label). For \textit{anchor selection}, we evaluate three alternatives: selecting (1) the window with minimum diversity, (2) a random window, and (3) the earliest window in the cluster. To ensure robustness, we repeated all experiments three times, and report mean F1 scores.

\begin{table}[htbp]
  \centering
  \vspace{-6pt}
  \caption{Ablation Study of Sampling Components}
  \vspace{-6pt}
\resizebox{0.38\textwidth}{!}{%
    \begin{tabular}{lccc}
    \toprule
          & \textbf{BGL} & \textbf{Hadoop} & \textbf{Thunderbird} \\
    \midrule
    Original \nm & \textbf{92.6}  & \textbf{97.3}  & \textbf{87.0} \\
    \hline
        - w. top-k token clustering & 88.4  & 95.6  & 82.1 \\
    \hline
        - w. random sampling & 90.4   & 97.3   & 80.1 \\
        - w. half random & 89.4   & 96.5   & 47.1 \\
        - w/o opposite sampling & 86.5   & 67.9   & 35.6 \\
    \hline
        - w. minimum diversity & 87.5  & 96.7  & 34.2 \\
        - w. random anchor & 90.2   & 92.7   & 69.5 \\
        - w. earliest anchor & 90.2   & 92.2   & 74.4 \\
    \bottomrule
    \end{tabular}%
}
  \vspace{-6pt}
  \label{tab:expr-rq2-sampling}%
\end{table}%

Table \ref{tab:expr-rq2-sampling} shows the F1 scores of different variants on the three datasets. 
When \nm's hierarchical window clustering is replaced by top-$k$ token clustering, the F1 score drops by 4.2\%/1.7\%/4.9\% on BGL/Hadoop/Thunderbird. The results indicate that iterative merging produces more meaningful clusters for contrastive sampling. 
All alternative sampling strategies result in decreased performance, with the most severe drop observed when contrastive sampling is removed (\eg F1 drops to 35.6\% on Thunderbird). This suggests that providing sufficient contrastive examples is critical for LLMs to accurately and stably learn discriminative rules. 
Additionally, for anchor selection, all three ablated variants yield lower F1 scores, especially on Hadoop and Thunderbird (\eg random anchor selection reduces 17.5\% F1 on Thunderbird). This highlights the importance of selecting representative anchors with diversity, which can reflect the distribution of each cluster and facilitate effective contrastive sampling.
These results show that our hierarchical window clustering and anchor-grounded window sampling are essential for constructing high-quality contrastive windows, thereby enabling LLMs for accurate recognition of normal and abnormal patterns.

\subsubsection{Configuration Analysis}\label{sec:expr-rq2-config}
\begin{figure}[t]
\captionsetup{justification=centering}
    \centering
    \includegraphics[width=0.4\textwidth]{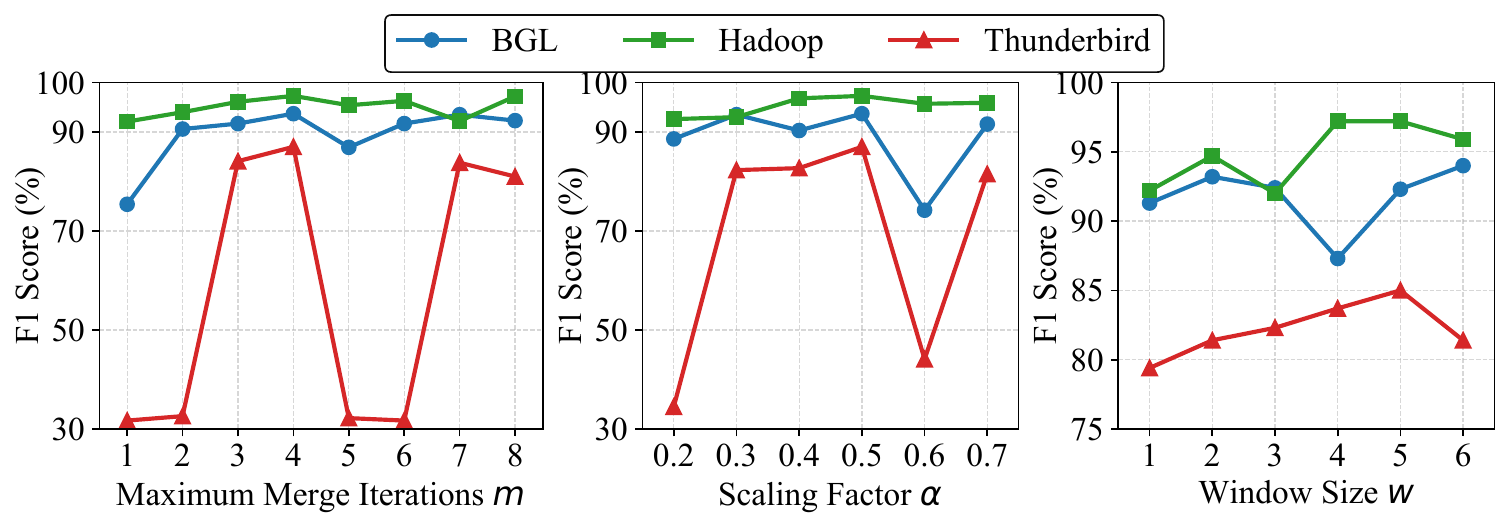}
    \caption{Configuration Analysis of \nm.}
    \label{fig:RQ2_ablation_cluster}
    \vspace{-10pt}
\end{figure}

Beyond the core clustering and sampling designs, we further investigate the impact of three key hyperparameters that could affect \nm's performance: (1) the maximum merge iteration $m$, (2) the scaling factor $\alpha$, and (3) the number of sampled contrastive windows $w$. These configurations influence the granularity of clusters and the diversity of sampled contrastive windows, which in turn affect the accuracy and generalizability of synthesized rules.
In this section, we explore different hyper-parameter settings for $m$, $\alpha$, and $w$ by conducting grid search. 
Specially, we vary one parameter at a time while keeping the others fixed at their default values ($m=4$, $\alpha=0.5$, $w=5$). The overall F1 scores under different settings are shown in Figure~\ref{fig:RQ2_ablation_cluster}.

\textbf{Maximum Merge Iteration ($m$):} 
The parameter $m$ determines the number of iterative merging steps in hierarchical clustering, directly influencing the granularity and coherence of clusters. As shown in the left subfigure of Figure~\ref{fig:RQ2_ablation_cluster}, increasing the maximum merge iteration size $m$ generally improves F1 scores on BGL and Hadoop, with optimal performance at $m=4$, which indicates that sufficient merging allows similar windows to be grouped together, facilitating effective contrastive sampling. 
When $m$ is set too low (\eg $m=1$), clustering becomes fragmented, resulting in scattered clusters that separate similar windows and hinder the formation of meaningful contrastive groups. This leads to significant drops in F1 scores (\eg $-18.3\%$ on BGL and $-5.2\%$ on Hadoop). 
On the Thunderbird dataset, performance is more sensitive to $m$, with large fluctuations observed. This suggests that improper clustering can result in poorly sampled contrastive pairs, making it difficult for rule synthesis. These results highlight the importance of choosing a proper merge iteration to balance cluster granularity and sample quality in \nm.

\textbf{Scaling Factor ($\alpha$):} The scaling factor $\alpha$ controls how many tokens represent each window during clustering, affecting the granularity of window features.
As shown in middle subfigure of Figure~\ref{fig:RQ2_ablation_cluster}, an intermediate value ($\alpha=0.5$) achieves the best F1 scores across datasets. 
This balanced setting ensures that enough representative tokens are used to capture the essential features of each window, facilitating effective clustering and sampling.
Too low $\alpha$ (\eg $0.2$) yields overly coarse representations, causing dissimilar windows to be grouped and leading to contrastive samples that mix different anomaly patterns. For example, we observe a drop in F1 of $5.1\%$/$4.7\%$/$52.5\%$ on BGL/Hadoop/Thunderbird, respectively. 
Conversely, a high value of $\alpha$ (\eg $0.7$) introduces noise into clustering, as more irrelevant tokens are included into window representations, making it harder to form coherent contrastive groups. As a result, the LLM may struggle to identify consistent rules, leading to F1 drops of $2.1\%$, $1.4\%$, and $5.5\%$ on the three datasets. These results indicate that a moderate scaling factor is crucial for balancing representation granularity and noise in \nm.

\textbf{Number of Sampled Windows ($w$):} 
The parameter $w$ controls the number of contrastive windows sampled for each anchor, directly affecting the diversity and informativeness of the training windows provided to the LLM. 
As shown in the right subfigure of Figure~\ref{fig:RQ2_ablation_cluster}, increasing $w$ from $1$ to $5$ leads to substantial improvements in F1 scores (\eg from $79.4\%$ to $85.0\%$ on Thunderbird and $92.2\%$ to $97.2\%$ on Hadoop). This is because a larger set of contrastive windows provides richer and more nuanced comparison clues, enabling the LLM to better distinguish subtle differences between normal and abnormal patterns.
However, further increasing $w$ beyond the optimal range results in a slight decline in performance. This can be attributed to the introduction of heterogeneous or noisy samples, as more windows are likely to contain diverse anomaly patterns or irrelevant information. The increased complexity makes it more challenging for the LLM to synthesize concise and generalizable rules, potentially leading to overfitting or reduced accuracy. These findings highlight the importance of selecting an appropriate number of contrastive samples in \nm.

\summary{\textbf{Answer to RQ2}: Hierarchical clustering and anchor-grounded sampling are essential for \nm’s accuracy; removing or altering these components leads to significant performance drops, indicating the importance of effective contrastive window collection for rule synthesis.}

\vspace{-6pt}
\subsection{RQ3: How do different agentic components affect \nm?}\label{sec:expr-rq3}
\subsubsection{Design Analysis on Synthesis Framework}\label{sec:expr-rq3-design}
In \nm, we employ an LLM-based multi-agent framework to synthesize executable rule code. To evaluate the contribution of each component in the agentic workflow, in this RQ, we first conduct an ablation study to evaluate the individual contribution of the components in the agentic workflow. To achieve this, we implement several variants of \nm by removing key synthesis modules in \nm, including, the refine agent, the repair agent, and the rollout module (\ie synthesizing only one rule in a single epoch). We repeated each experiment three times to mitigate randomness, and report mean scores.

\begin{table}[htbp]
  \centering
  \vspace{-8pt}
  \caption{Ablation Study of Synthesis Components}
  \vspace{-6pt}
\resizebox{0.3\textwidth}{!}{
    \begin{tabular}{lccc}
    \toprule
          & \textbf{BGL} & \textbf{Hadoop} & \textbf{Thunderbird} \\
    \midrule
    \nm & 92.6  & 97.3  & 87.0 \\
    - w/o refine agent & 89.1  & 93.1  & 82.7 \\
    - w/o repair agent & 85.8  & 97.0  & 30.4 \\
    - w/o rollout & 87.7  & 95.4  & 83.7 \\
    \bottomrule
    \end{tabular}%
}
  \vspace{-8pt}
  \label{tab:expr-rq3-synthesis}%
\end{table}%

Table~\ref{tab:expr-rq3-synthesis} shows the F1 scores of different variants across the three datasets. Removing any module—refine agent, repair agent, or rollout—results in a notable performance drop, indicating the importance of each component in the agentic synthesis workflow. 
Excluding the refine agent leads to F1 drops of 3.5\%, 4.2\%, and 4.3\% on BGL, Hadoop, and Thunderbird, respectively. This highlights the refine agent's role in improving rule generalizability and accuracy by iteratively enhancing the synthesized code.
Specially, we found 46.4\%/100.0\%/55.1\% of the rules are improved by the refine agent to cover (correctly predict) more windows across the three datasets. 
The impact of removing the repair agent varies by dataset complexity. On Hadoop, which contains fewer log messages and templates~\cite{jiang2023large}, the performance drop is marginal (from 97.3\% to 97.0\%). In contrast, on the more complex Thunderbird dataset, the F1 score plummets from 87.0\% to 30.4\%, indicating that the repair agent is crucial for handling challenging log patterns and ensuring syntactic correctness.
Omitting the rollout module results in F1 decreases of 4.9\%, 1.9\%, and 3.3\% on BGL, Hadoop, and Thunderbird, respectively. This demonstrates the benefit of iterative rule synthesis and selection, which enables the framework to discover more robust and generalizable rules.
Overall, these results confirm that each agentic component contributes to the effectiveness and robustness of the \nm synthesis framework.

\subsubsection{Different LLMs}\label{sec:expr-rq3-llm}
In our experiments, \textit{gpt-5-mini} serves as the default LLM for rule synthesis in \nm. In this RQ, we further assess the adaptability and robustness of \nm evaluating its performance with a diverse set of LLMs. 
Specifically, we test eight popular closed-source LLMs of varying sizes, including GPT-5 series (\textit{gpt-5}, \textit{gpt-5-mini}, \textit{gpt-5-nano})~\cite{gpt5}, GPT-4.1 series (\textit{gpt-4.1-20250414}, \textit{gpt-4.1-mini-20250414}, \textit{gpt-4.1-nano-20250414})~\cite{gpt41}, and GPT-4o (\textit{gpt-4o-20241120}, \textit{gpt-4o-mini})\cite{gpt4o}.
as well as six leading open-source LLMs specialized for coding tasks, \ie GPT-oss (\textit{gpt-oss-20b}, \textit{gpt-oss-120b})\cite{openai2025gptoss120bgptoss20bmodel}, Qwen3-Coder (\textit{Qwen3-Coder-480B-A35B-Instruct})~\cite{qwen3technicalreport}, Qwen3-235B (\textit{Qwen3-235B-A22B-Instruct-2507})~\cite{qwen3technicalreport} DeepSeek-V3.1 (\textit{DeepSeek-V3.1-250821})~\cite{deepseekai2024deepseekv3technicalreport} and DeepSeek-R1 (\textit{DeepSeek-R1-250528})~\cite{guo2025deepseekr1}.
Table~\ref{tab:expr-rq3-llm-f1only} reports the precision, recall, and F1 scores of \nm on the three datasets for each LLM.
Across all LLMs, \nm demonstrates consistently high performance, with average F1 scores exceeding 88\% for most models.

\begin{table}[htbp]
  \centering
  \vspace{-10pt}
  \caption{Ablation study of different LLMs in \nm (\%)}
  \vspace{-6pt}
\resizebox{0.4\textwidth}{!}{%
  \begin{tabular}{lcccc}
    \toprule
    \textbf{LLM used in \nm} & \textbf{BGL} & \textbf{Hadoop} & \textbf{Thunderbird} & \textbf{Avg.} \\
    \midrule
    gpt-5 & \textbf{97.7} & 96.5 & 82.1 & 92.1 \\
    gpt-5-mini & 92.6 & \textbf{97.3} & \textbf{87.0} & \textbf{92.3} \\
    gpt-5-nano & 92.1 & 96.4 & 78.1 & 88.9 \\
    gpt-4.1-20250414 & 94.1 & 95.6 & 83.2 & 91.0 \\
    gpt-4.1-mini-20250414 & 93.8 & 96.3 & 79.6 & 89.9 \\
    gpt-4.1-nano-20250414 & 90.0 & 80.6 & 83.4 & 84.6 \\
    gpt-4o-20241120 & 91.1 & 93.2 & 87.0 & 90.4 \\
    gpt-4o-mini & 93.7 & 83.4 & 81.6 & 71.7 \\
    \midrule
    gpt-oss-120b & 94.5 & 96.2 & 79.2 & 90.0 \\
    gpt-oss-20b & 91.2 & 95.7 & 81.1 & 89.3 \\
    Qwen3-Coder-480B-A35B-Instruct & 76.6 & 96.4 & 81.9 & 84.9 \\
    Qwen3-235B-A22B-Instruct-2507 & 93.7 & 85.3 & 31.5 & 70.2 \\
    DeepSeek-V3.1-250821 & 85.6 & 95.1 & 70.5 & 83.7 \\
    DeepSeek-R1-250528 & 95.8 & 95.5 & 47.6 & 79.6 \\
    \bottomrule
  \end{tabular}
}
  \vspace{-4pt}
  \label{tab:expr-rq3-llm-f1only}
\end{table}

\noindent\textbf{Closed-source LLMs:} 
For closed-source LLMs, we observe that newer versions exhibit  advantages over their predecessors. For example, the average F1 scores are 92.1\% for \textit{gpt-5}, 91.0\% for \textit{gpt-4.1-20250414}, and 90.4\% for \textit{gpt-4o-20241120}. We attribute the gains to the enhanced coding and reasoning capabilities in newer models~\cite{gpt5}, which are beneficial for complex anomaly pattern inference and rule synthesis.

\noindent\textbf{Open-source LLMs:} 
Among open-source LLMs, \nm with \textit{gpt-oss-120b} achieves an average F1 of 90.0\%, closely matching the best closed-source models. 
Other open-source LLMs, such as DeepSeek-V3.1 and Qwen3-Coder, also show competitive results, further demonstrating the robustness of \nm when adapted to diverse LLMs. Notably, the strong performance of large open-source models enables the possibility of local deployment, which is critical for scenarios involving sensitive or proprietary log data.

\noindent\textbf{Model Size Analysis:} 
Interestingly, larger models do not always outperform smaller ones, especially for the newest LLMs. For instance, \textit{gpt-5-mini} achieves a higher average F1 (92.3\%) than \textit{gpt-5} (91.1\%) and \textit{gpt-5-nano} (88.9\%). 
Similarly, \textit{gpt-oss-20b} (89.3\%) performs comparably to \textit{gpt-oss-120b} (90.0\%) and even surpasses it on the Thunderbird dataset (81.1\% vs. 79.2\% F1).
We attribute steady performance across model sizes to the contrastive window sampling modules and agentic synthesis scaffold in \nm, which provide informative, high-quality contrastive log windows for rule inference. Since the synthesized code is relatively simple and self-contained, smaller models are sufficient for this task, unlike more complex code generation tasks that require a longer context and output~\cite{wang2024rlcoder, huang2024contextualized, hou2024llm4sesurvey}. These findings suggest that LLM-based rule synthesis may not strictly follow the scaling law~\cite{kaplan2020scaling} for \task, supporting the use of smaller, more cost-effective models in practical deployments.

\summary{\textbf{Answer to RQ3}: All agentic components—refine, repair, and rollout—are crucial for rule synthesis. \nm maintains strong performance across diverse LLMs including smaller and open-source LLMs, which suggest that practitioners can flexibly choose LLMs to balance cost and accuracy.}

\subsection{RQ4: How efficient and cost-effective is \nm in monitoring large-scale log data?}\label{sec:expr-rq4}

\subsubsection{Efficiency Analysis on Online Inference}\label{sec:expr-rq4-time}
Efficiency is critical for log-based anomaly detection in real-world scenarios due to the massive volume of logs generated~\cite{ma2025practitioners, he2022empiricalMicrosoft}. In this RQ, we evaluate the online inference efficiency of \nm compared to eight baseline methods, including rule-based, ML-based, DL-based, and LLM-based approaches. We measure the total execution time for each method on raw streaming logs, including any required preprocessing steps (\eg grouping and parsing), to simulate realistic deployment. We conducted all experiments  on the same machine without GPU acceleration to ensure fair comparison.

\begin{figure}[t]
    \centering
    \includegraphics[width=0.48\textwidth]{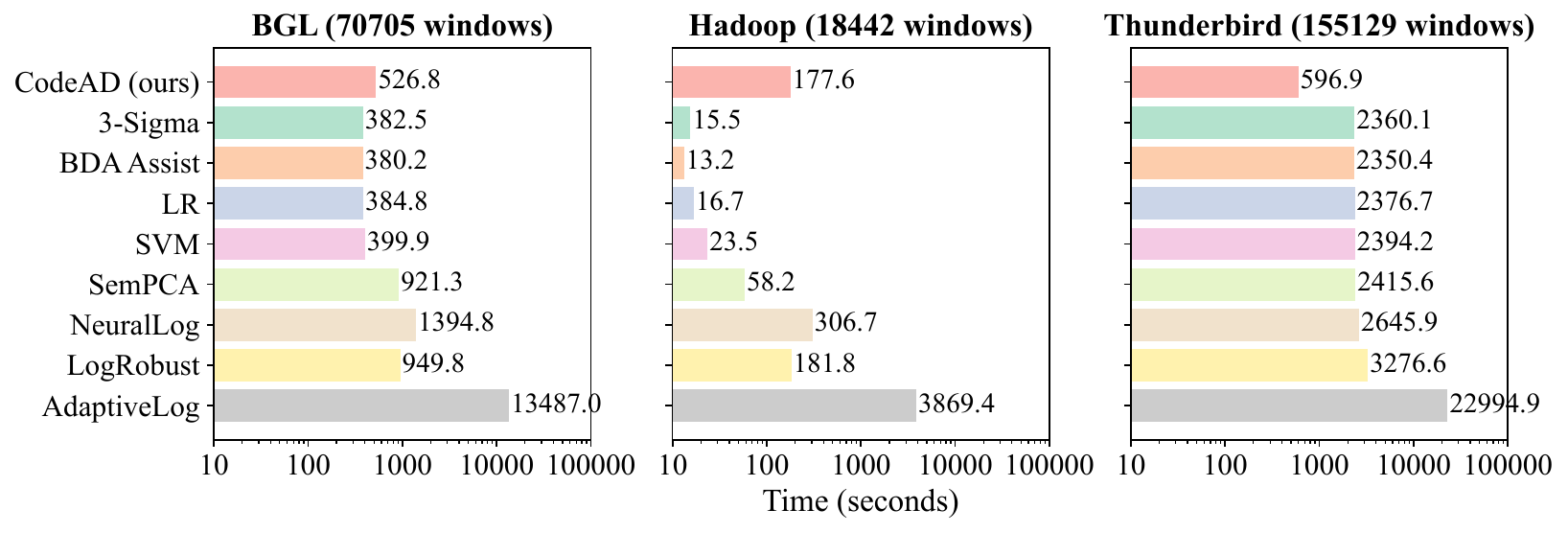}
    \vspace{-10pt}
    \caption{Real-time efficiency of \nm and baselines (second).}
    \label{fig:RQ4_efficiency}
    \vspace{-10pt}
\end{figure}

\textbf{Results:} Figure~\ref{fig:RQ4_efficiency} shows the execution time and number of windows for each test dataset. We can observe that \nm consistently achieves higher online efficiency than DL-based and LLM-based baselines. For example, compared to AdaptiveLog—the most accurate baseline—\nm achieves superior detection accuracy with only 3.2\% of the average inference time across all datasets. 
When compared to rule-based and ML-based methods, \nm's efficiency varies with dataset size. On smaller datasets (\eg Hadoop with 18,442 windows), \nm is slower than the fastest rule-based method (177.6s for \nm vs. 13.2s for BGD Assist). However, as dataset size increases, \nm becomes more competitive: on BGL (70,705 windows), \nm and BGD Assist have comparable inference times (526.8s vs. 380.2s), and on the largest dataset, Thunderbird (155,129 windows), \nm outperforms all baselines (596.9s vs. 2,350.4s for BGD Assist). 
The efficiency gap is primarily due to the preprocessing overhead in rule-based and ML-based methods, especially log parsing, which becomes a bottleneck on larger and more complex datasets (\eg parsing costs 2,340.6s on Thunderbird with Drain~\cite{he2017drain})~\cite{jiang2023large}. 
In contrast, \nm does not require log parsing or specialized hardware, making it highly practical for real-time industrial applications where log volume and complexity are substantial.

\subsubsection{Cost Analysis of Offline Synthesis}\label{sec:expr-rq4-offline-cost}
\nm leverages LLMs for rule generation, repair, and refinement in its agentic synthesis workflow. Unlike traditional model-based methods, which can be trained efficiently on local hardware, LLM-based approaches typically rely on cloud-based inference services due to the substantial computational resources required for LLMs. These services, such as those provided by OpenAI, charge users based on the number of input and output tokens processed during each API call, reflecting the underlying computational and infrastructure costs. 
As a result, the financial cost of using LLMs is directly tied to the token consumption, making it a key factor in evaluating practical feasibility. To assess this, we analyze the financial cost of \nm using different LLMs, specifically comparing \textit{gpt-5.1} and \textit{gpt-5.1-mini}. We report F1 scores, total token consumption, and average token usage per iteration, averaged across all datasets.

\begin{table}[htbp]
  \centering
  \vspace{-8pt}
  \caption{Cost Analysis}
  \vspace{-8pt}
\resizebox{0.3\textwidth}{!}{
    \begin{tabular}{|l|r|r|}
    \hline
          & \multicolumn{1}{l|}{\textit{gpt-5}} & \multicolumn{1}{l|}{\textit{gpt-5-mini}} \\
    \hline
    Avg. F1 & 92.1\% & 92.3\% \\
    \hline
    \# epoch & 130.3 & 65.7 \\
    \# input tokens & 2.7M & 2.5M \\
    \# output tokens & 3.5M & 1.7M \\
    \$ total cost & \$38.55 & \$3.99 \\
    \$ cost per rule & \$0.32 & \$0.07 \\
    \hline
    \$ generation agent & \$24.62 & \$3.37 \\
    \$ repair agent & \$0.15 & \$0.03 \\
    \$ refine agent & \$13.78 & \$0.60 \\
    \hline
    \end{tabular}%
}
  \vspace{-6pt}
  \label{tab:expr-rq4-cost}%
\end{table}%

Table~\ref{tab:expr-rq4-cost} summarizes the average cost for each dataset for \nm with different LLMs. We observe that: 
(1) \nm with \textit{gpt-5.1-mini} achieves higher detection accuracy (+0.2\% F1 on average) while reducing synthesis costs to \$0.07 per rule and \$3.99 in total, demonstrating that accurate rule generation can be achieved without excessive financial overhead.
(2) Among the agentic components, generation agents incur the highest cost (\$24.62/\$3.37 for \textit{gpt-5}/\textit{gpt-5-mini}), while repair agents are the least expensive (\$0.15/\$0.02 for \textit{gpt-5}/\textit{gpt-5-mini}), which is likely due to the high success rate of LLMs in generating syntax-correct code. Refine agents incur moderate costs (\$13.78/\$0.60 for \textit{gpt-5}/\textit{gpt-5-mini}), reflecting the need of additional refinement to improve rule generalizability.
Overall, these results demonstrate that \nm provides a cost-effective and efficient solution for \task. By balancing cost and accuracy, \nm provides a practical alternative for practitioners seeking high-performance methods with minimal financial overhead.

\summary{\textbf{Answer to RQ4}: \nm is efficient and cost-effective for \task, processing large-scale logs faster than baselines and requiring minimal construction expenses, making it practical for scalable, real-time monitoring in production.}

\subsection{RQ5: How can \nm benefit engineers?}\label{sec:expr-rq5}

\noindent To intuitively understand practical benefits of \nm for engineers, we characterize the interpretability and complexity of synthesized rules through case studies and statistical analysis.

\begin{figure}[h]
    \centering
    \includegraphics[width=0.9\columnwidth]{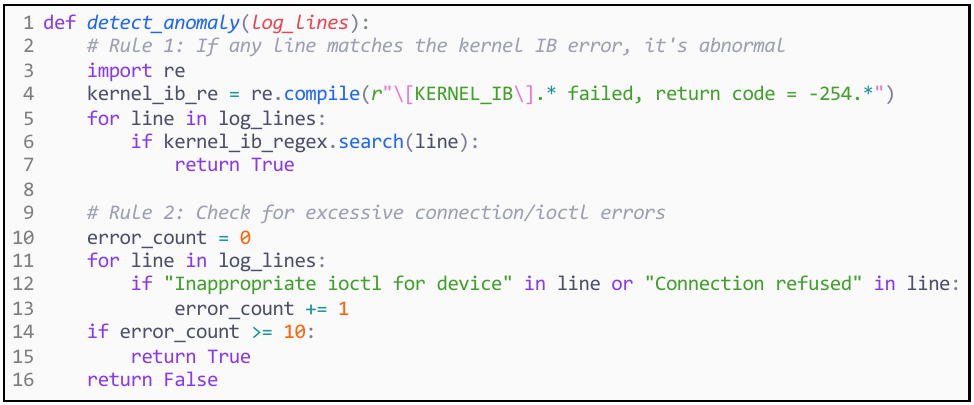}
    \caption{Case study: a rule code function synthesized by \nm.}
    \label{fig:case-study}
    \vspace{-6pt}
\end{figure}

\subsubsection{Case Study}
Figure~\ref{fig:case-study} shows a representative anomaly detection function synthesized by \nm from the Thunderbird dataset. As shown in the example, the synthesized function rule comprises two subrules, enabling detection of different aspects of anomalies. 
The first subrule is a keyword matching type, which checks the existence of ``KERNAL\_IB [\dots] failed, return code = -254 [\dots]'' in log windows. 
The second subrule counts for the log event occurrence, where more than 10 appearance of connection or ioctl errors in log windows will be judged as abnormal. 
Before each subrule, clear comments to specify the subrule are included, which can enhance transparency and facilitate human understanding.

\subsubsection{Complexity Analysis}
Due to the complexity of anomalies, each synthesized rule (\ie a function) can contain multiple subrules to detect different aspects of the anomaly. To characterize subrule types, we used GPT-4.1 for automated classification. 
For the subrule categories, we reuse the seven types listed in Table~\ref{tab:background-rules} and add one more ``Other'' category to detect outliers as the total. To ensure the classification accuracy, we manually annotate a random testset of 50 subrules and found GPT-4.1 achieve an accuracy of 94\%. 

\begin{figure}[h]
    \centering
    \includegraphics[width=0.9\columnwidth]{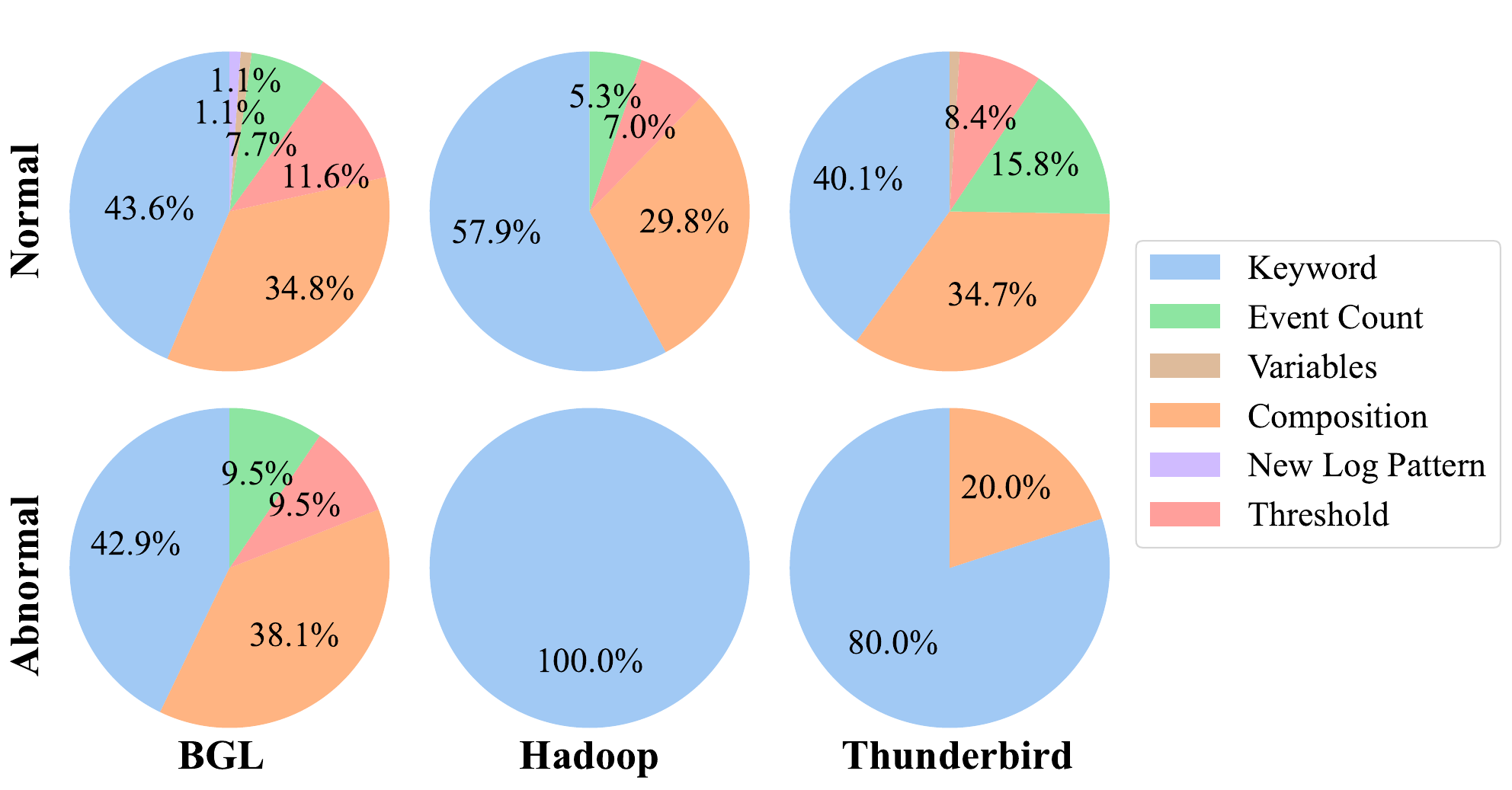}
    \vspace{-8pt}
    \caption{Distribution of subrules across different datasets.}
    \label{fig:RQ5_subrule_distribution}
    \vspace{-2pt}
\end{figure}

\textbf{Subrule Type Distribution.}
Figure~\ref{fig:RQ5_subrule_distribution} shows the distribution of subrule types generated from each dataset. We find that, Keyword matching is the most prevalent type, accounting for 38.9\% (BGL), 50.8\% (Hadoop), and 35.1\% (Thunderbird) of normal rules, and up to 100\% of abnormal rules in Hadoop. This indicates that many anomalies can be detected using simple keyword matching, which is consistent with previous findings~\cite{yu2024DLorML, ali2025mlLOD}. Composition is another frequent subrule type, accounting for 31.0\% (BGL), 26.2\%(Hadoop), 30.3\% (Thunderbird) of normal rules, and 36.4\% (BGL) for abnormal rules. This indicates that normality and anomalies are still complex, which can combine multiple subrules with ``AND'' and ``OR'' conditions. 
Other common subrule types include threshold checks, event counts, and variable analysis. These results demonstrates \nm's ability to synthesize diverse and complex rules and have the potentials to also perform well on more complex datasets in real-world systems.

\begin{table}[htbp]
\centering
\small
\caption{Statistics of synthesized rules}
\resizebox{0.38\textwidth}{!}{%
\begin{tabular}{|lccc|}
    \hline
        & \textbf{BGL} & \textbf{Hadoop} & \textbf{Thunderbird} \\
        \hline
        \# normal rules & 68 & 23 & 80 \\
        \# abnormal rules & 15 & 5 & 6 \\
        \hline
        \multicolumn{4}{|l|}{\textit{Normal rules (avg. per rule)}} \\
        \# code lines & 51.57 & 52.61 & 53.44 \\
        cyclomatic complexity & 8.2 & 6.4 & 7.6 \\
        \# statements & 12.0 & 8.8 & 11.1 \\
        \# sub rules & 3.0 & 2.8 & 2.9 \\
        \hline
        \multicolumn{4}{|l|}{\textit{Abnormal rules (avg. per rule)}} \\
        \# code lines & 14.9 & 9.6 & 14.0 \\
        cyclomatic complexity & 3.7 & 3.0 & 3.5 \\
        \# statements & 5.1 & 4.0 & 4.7 \\
        \# sub rules & 1.5 & 1.0 & 1.2 \\
    \hline
\end{tabular}
}
\label{tab:expr-rq5-statistics}
\end{table}

\textbf{Rule Complexity Statistics.}
Table~\ref{tab:expr-rq5-statistics} summarizes the complexity of synthesized rules across three datasets. We find that normal rules are substantially more numerous than abnormal rules (\eg 68 vs. 15 in BGL, 23 vs. 5 in Hadoop, and 80 vs. 6 in Thunderbird), reflecting the predominance of normal system behavior in operational logs. Normal rules also exhibit greater complexity. On average, each normal rule contains 51.6/52.6/53.4 lines of code in BGL/Hadoop/Thunderbird, respectively, compared to only 14.9/9.6/14.0 lines for abnormal rules. Similarly, normal rules have higher cyclomatic complexity (8.2/6.4/7.6 vs. 3.7/3.0/3.5) and more statements (12.0/8.8/11.1 vs. 5.1/4.0/4.7). The average number of subrules per normal rule is also higher (3.0/2.8/2.9) than for abnormal rules (1.5/1.0/1.2). These findings suggest that identifying normal behavior requires more elaborate logic and rule composition, likely due to the diversity and complexity of normal operational patterns. For engineers, this means that maintaining and reviewing normal rules may demand greater effort, but also provides more comprehensive coverage of system states. In contrast, abnormal rules tend to be simpler and more focused, facilitating rapid anomaly detection.

The results demonstrate that \nm produces interpretable, well-documented rules that support transparency and facilitate human review. The diversity of synthesized subrules enables comprehensive anomaly detection, while the clear structure and comments enhance maintainability and ease of integration into engineering workflows.

\summary{\textbf{Answer to RQ5}: \nm generates interpretable, well-structured Python rules that enhance transparenc, facilitating easier integration and trust in automated \task.}

\section{Discussion}

\subsection{Practical Usage Scenarios}

In this section, we discuss potential applications of \nm and outline  future directions. By automating the synthesis of interpretable anomaly detection rules from log data, \nm can significantly enhance the efficiency and effectiveness of Site Reliability Engineers (SREs) in large-scale system operations. Below, we highlight two primary usage scenarios.

\noindent\textbf{Automatic Alert Recommendation.}
A common challenge in modern operations is the rapid configuration of effective alerts following new incidents. Traditionally, after an incident, SREs must manually analyze logs, identify anomaly patterns, and encode new detection rules—a process that is both time-consuming and error-prone. With \nm, this workflow can be streamlined: after an incident is investigated and relevant logs are collected (e.g., from the affected component), \nm can automatically mine anomaly patterns and synthesize executable, human-readable rule code. These rules can be directly integrated into existing monitoring pipelines to enrich alert coverage. Furthermore, the generated code serves as a transparent interface for engineers, who can review, customize, or retire rules as systems evolve. This approach not only accelerates the deployment of new alerts, but also ensures that alert logic remains interpretable and maintainable.

\noindent\textbf{Root Cause Analysis and Log Reduction.}
Timely root cause analysis is critical for minimizing incident impact; even one hour of delay can lead to significant losses. However, the sheer volume and complexity of logs in distributed systems often overwhelm engineers. \nm can assist by filtering out irrelevant log windows and highlighting interpretable clues. For example, \nm maintains a library of normal and abnormal rules: normal rules can be used to quickly exclude normal log windows, while abnormal rules help identify anomaly patterns. Log windows that do not match any existing rules can be prioritized for manual investigation. This targeted approach reduces the volume of logs that need review, allowing SREs to focus on the most relevant data and accelerating the diagnosis process. In practice, such log reduction and prioritization can significantly improve incident response times and reduce operational overhead.

\subsection{Threats To Validity}

\noindent\textbf{Randomness}
Randomness in LLM-based synthesis can arise from stochastic sampling, model temperature, and agentic rollouts. To minimize this threat, we set the LLM temperature to zero for deterministic outputs (if the LLM has the temperature hyperparameter) and conducted each experiment three times under every setting, reporting the average results. This approach reduces the impact of random fluctuations and ensures the robustness of our findings.

\noindent\textbf{Implementation and settings}
Implementation bias may arise from differences in codebases, parameters, or experiment environments. To address this, we adopted official replication packages and benchmark for all baselines, using default parameters and settings from prior work~\cite{he2016experience}. Where necessary, we re-implemented baselines following published paper and conducted peer code reviews to ensure correctness. We also validated our results against recent benchmarks to confirm consistency.

\noindent\textbf{Dataset Representativeness and Generalizability.}
Our evaluation is conducted on datasets collected from large-scale software systems, which may not cover all possible log formats or operational environments. While these datasets are representative of modern cloud operations, there remains a risk that \nm may not generalize to other domains or log types. Future work will involve testing \nm on a broader range of datasets, including open-source and industrial logs from different platforms, to further assess its generalizability.

\section{Related Work}

\subsection{Log-based Anomaly Detection}

Log-based anomaly detection is promising for enhancing system reliability~\cite{he2021logsurvey, soldani2022ADRCAsurvey}, which is important in various domains~\cite{yu2025cashift, wang2024exploratory}. Accurate detection can benefit thereafter bug localization~\cite{chen2021pathidea}, fault diagnosis~\cite{amar2019mining, wang2020LogadForRCA, zhang2022deeptralog} and issue resolution~\cite{mahindru2021logadForResolution}. 
The earliest studies in log-based anomaly detection involved constructing detectors via machine-learning approaches~\cite{he2017towardsPCA, bodik2010fingerprintingLR, liu2008isolationIR, liang2007ibmSVM, lin2016logcluster, chen2004decisiontree}, which mainly employ classification or clustering paradigms to build efficient detectors.
However, these methods rely on accurate parsing techniques to obtain log templates~\cite{le2021neurallog, khan2023impact, shin2021theoretical} and can only recognize simple abnormal patterns~\cite{zhang2024automl}, results in unsatisfactory performance.
To achieve more accurate anomaly detection, researchers have developed deep learning-based methods for log-based anomaly detection ~\cite{du2017deeplog, meng2019loganomaly, zhang2022cat, guo2021logbert, yang2021semiPLELog, le2021neurallog, zhang2019Logrobust}. Based on system event frequencies or the semantics of log entries, these methods map log sequences to numerical vectors and then train a neural network to classify the sequences for detecting system anomalies. 
Although highly effective~\cite{huang2020hitanomaly, zhang2022cat, guo2024logformer, chu2025anomaly, tang2024substructure, he2025weakly}, these methods can only output a single detection result and, by using neural networks for detection, they perform poorly in terms of interpretability and efficiency.
A recent research hotspot is the use of large language models for log-based anomaly detection~\cite{ji2024adapting, xu2025gelog, li2025anomalygen, guan2024logllm, ma2025adaptivelog, he2024llmelog}, with these approaches aim to leverage the LLM's understanding of system log semantics for anomaly detection through techniques such as prompting~\cite{qi2023loggpt, liu2024logprompt, liu2024loglm, liu2023sealog, huang2025logrules}, RAG~\cite{duan2025eagerlog, pan2024raglog, zhang2024ragLogAD}, and model collaboration~\cite{ma2025adaptivelog, xiao2025clslog}. Although they have improved effectiveness and interpretability, their detection efficiency is unacceptable due to the massive parameter scale of LLMs. 
\nm distinguishes from these methods by synthesizing anomaly detection rules in terms of executable Python code with LLMs, which supports fast and accurate online detection while providing explainability. 
Among these methods, LogRules~\cite{huang2025logrules} applies a similar LLM-based rule extraction paradigm to \nm, which first employs LLMs to produce natural language rules and then prompts a smaller-scale LLM (\eg LLaMA-3-8B-Instruct) with these rules to infer labels for windows. However, \nm synthesizes executable Python code rules, which is efficient and cost-effective.

\subsection{LLM for Log Analysis}

Log analysis is long-standing topic in software engineering. Logs record the system's operational state, we can obtain system runtime information for various software development and maintenance tasks by analyzing logs~\cite{he2021logsurvey}. 
With the rise of large language models, researchers have in recent years explored their capabilities for various log analysis tasks~\cite{zhang2025aiopssurvey}, including log parsing, root cause localization (RCA), and logging statement generation.
The objective of log parsing is to convert unstructured raw logs into structured log events. Most recent work~\cite{jiang2023llmparser,ma2024llmparser, xiao2024free, xu2024divlog, wang2025inferlog} specializes LLMs for the log parsing task by fine-tuning or in-context learning on labeled data, achieving notable performance. Other study~\cite{huang2025lunar} has also explored the potential of LLMs for log parsing in unannotated scenarios by extracting log contrastive units. 
Logging statement generation aims to automatically produce log statements, creating logs that record system behavior and aid software maintenance. Recent works~\cite{li2024go, li2023exploring} primarily employ in-context learning, leveraging LLMs with very few examples to generate high-quality logging statements. Other work~\cite{zhong2025logupdater} fine-tunes LLMs to generate logs that meet unique organizational requirements, such as specific formats or compliance standards. 
The objective of the RCA task is to identify the root cause of failures based on the system's runtime logs. Recent work~\cite{xu2025openrca, li2025coca} primarily utilizes LLMs to construct agent systems that analyze collected system runtime logs by integrating well-designed data analysis tools or code knowledge to report the root cause. 
Another hotspot for using LLMs in log analysis is fault information extraction, which aims to retrieve failure-related information from runtime logs for further analysis by SREs. These efforts~\cite{huang2024lofi, wang2024large} are primarily based on in-context learning, using carefully crafted exemplars to prompt LLMs to extract fine-grained log information that may be of interest to engineers. 
In this paper, we aim to fully leverage LLMs' comprehension of log semantics and software system anomalies to summarize system anomaly patterns from massive runtime logs and describe them in the form of executable code, enabling accurate, efficient, and interpretable system anomaly detection.

\section{Conclusion}

In this work, we propose \nm, a novel framework that automatically synthesizes executable Python rule functions for log-based anomaly detection using LLMs. By integrating hierarchical clustering, anchor-grounded sampling, and an agentic synthesis workflow, \nm generates efficient, interpretable, and generalizable detection rules, reducing manual effort and improving scalability. 
Experiments on public datasets demonstrate that \nm consistently outperforms state-of-the-art baselines, while maintaining superior inference efficiency and low synthesis cost. Our ablation studies further confirm the effectiveness of each core component, and analysis of the synthesized rules highlights their transparency and practical value for engineers. 
These results highlight the potential of LLM-driven rule synthesis to bridge the gap between automation, interpretability, and efficiency in real-world anomaly detection.

\section*{Acknowledgments}
The work described in this paper was supported by the Research Grants Council of the Hong Kong Special Administrative Region, China (No. SRFS24254S03 of the Senior Research Fellow Scheme and No. CUHK 14209124 of the General Research Fund). Domenico Bianculli was supported by the Luxembourg National Research Fund (FNR), under grant reference C22/IS/17373407/LOGODOR.

\balance

\bibliographystyle{IEEEtran}
\bibliography{reference}

\end{document}